\documentclass[12pt]{article}
\usepackage{epsfig, amsmath, amssymb}
\usepackage{hyperref}
\usepackage{graphicx,psfrag}
\usepackage{xcolor}
\newcommand{\nn}{\nonumber}
\newcommand{\be}{\begin{equation}}
\newcommand{\ee}{\end{equation}}
\newcommand{\ben}{\begin{equation}}
\newcommand{\een}{\end{equation}}
\newcommand{\bea}{\begin{eqnarray}}
\newcommand{\eea}{\end{eqnarray}}
\newcommand{\bA}{\begin{array}}
\newcommand{\eA}{\end{array}}
\newcommand{\bc}{\begin{center}}
\newcommand{\ec}{\end{center}}
\newcommand{\al}{\alpha}

\newcommand{\ra}{\rightarrow}
\newcommand{\del}{\partial}

\newcommand{\ie}{{\it i.e.}}
\newcommand{\eg}{{\it e.g.}}

\newcommand{\lan}{\langle}
\newcommand{\ran}{\rangle}

\begin{document}


\begin{titlepage}

%

\bc

\hfill 
\\         [15mm]

{\Huge No-boundary extremal surfaces in 
 \\ [2mm]
   slow-roll inflation and other cosmologies} 
\vspace{16mm}

{\large Kaberi Goswami,\ \ K.~Narayan,\ \ Gopal Yadav} \\
\vspace{3mm}
{\small \it Chennai Mathematical Institute, \\}
{\small \it H1 SIPCOT IT Park, Siruseri 603103, India.\\}

\ec
\vspace{35mm}

\begin{abstract}
  Building on previous work on de Sitter extremal surfaces anchored at
  the future boundary, we study no-boundary extremal surfaces in
  slow-roll inflation models, with perturbations to no-boundary global
  $dS$ preserving the spatial isometry. While in pure de Sitter space
  the Euclidean hemisphere gives a real area equalling half de Sitter
  entropy, the no-boundary extremal surface areas here have nontrivial
  real and imaginary pieces overall. We evaluate the area integrals in
  the complex time-plane defining appropriate contours. For the 4-dim
  case, the real and imaginary finite corrections at leading order in
  the slow-roll parameter match those in the semiclassical expansion
  of the Wavefunction (or action), and corroborate the cosmic brane
  interpretation discussed previously. We also study no-boundary
  extremal surfaces in other cosmologies including 3-dimensional
  inflation and Schwarzschild de Sitter spaces with small mass.
\end{abstract}


\end{titlepage}

{\tiny 
\begin{tableofcontents}    
\end{tableofcontents}
}



\section{Introduction}

It is of great interest to understand cosmology from the point of view
of holography \cite{Maldacena:1997re,Gubser:1998bc,Witten:1998qj}. One
of the many fascinating aspects of our understanding of holography in
$AdS$-like spaces is the Ryu-Takayanagi formulation of holographic
entanglement entropy
\cite{Ryu:2006bv,Ryu:2006ef,HRT,Rangamani:2016dms} and its many
ramifications.  Towards exploring these quantum information ideas in
cosmological contexts, it is fascinating to look at de Sitter space as
a idealized example of cosmology. This however already has many new
features. Thinking of time as the holographic direction, a natural
boundary for de Sitter space is in the far future (or past), leading
to $dS/CFT$
\cite{Strominger:2001pn,Witten:2001kn,Maldacena:2002vr,Anninos:2011ui}\
(see \eg\ \cite{Spradlin:2001pw,Anninos:2012qw,Galante:2023uyf} for
reviews on de Sitter space and its entropy \cite{Gibbons:1977mu}).

Certain generalizations of the Ryu-Takayanagi formulation of $AdS$
holographic entanglement to de Sitter space were studied in
\cite{Narayan:2015vda,Narayan:2015oka,Sato:2015tta,Miyaji:2015yva,
  Narayan:2017xca,Narayan:2019pjl,Narayan:2020nsc,Doi:2022iyj,Narayan:2022afv,Narayan:2023zen}.
These pertain to RT/HRT extremal surfaces anchored at the $dS$ future
boundary $I^+$ and amount to considering the bulk analog of setting up
entanglement entropy in the dual CFT at $I^+$, in part towards
understanding if de Sitter entropy \cite{Gibbons:1977mu} can be
understood as some sort of holographic entanglement entropy. Analysing
the extremization shows that surfaces anchored at $I^+$ do not return
to $I^+$. In entirely Lorentzian $dS$, this leads to future-past
timelike surfaces stretching between $I^\pm$ with pure imaginary
area. With a no-boundary type Hartle-Hawking boundary condition, the
top half of these timelike surfaces joins with a spacelike part on the
hemisphere giving a complex-valued area \cite{Doi:2022iyj},
\cite{Narayan:2022afv}\ (and \cite{Hikida:2022ltr,Hikida:2021ese} for
$dS_3/CFT_2$). The real part of the area arises from the hemisphere
and is precisely half de Sitter entropy (Figure~\ref{fig1}). Some of
these structures
\cite{Narayan:2017xca,Narayan:2019pjl,Narayan:2020nsc} are akin to
space-time rotations from $AdS$ (some related discussions appear in
\cite{Cotler:2023xku}). In \cite{Narayan:2023zen}, certain analytic
continuations were discussed which in fact amount to space-time
rotations from $AdS$.

Recent investigations suggest that the areas of these extremal
surfaces are best interpreted as encoding pseudo-entropy or
time-entanglement \cite{Doi:2022iyj}, \cite{Narayan:2022afv},
entanglement-like structures involving timelike
separations. Pseudo-entropy \cite{Nakata:2020luh} is the entropy based
on the transition matrix $|f\ran\lan i|$ regarded as a generalized
density operator (see also \cite{Li:2022tsv}-\cite{Heller:2024whi} for
some partially related work). In some sense this is perhaps the
natural object here since the absence of $I^+\ra I^+$ returns for
extremal surfaces suggests that extra data is required in the
interior, somewhat reminiscent of scattering amplitudes (equivalently
the time evolution operator), and of \cite{Witten:2001kn} viewing de
Sitter space as a collection of past-future amplitudes. This is also
suggested by the $dS/CFT$ dictionary
$Z_{CFT}=\Psi_{dS}$\ \cite{Maldacena:2002vr}: boundary ``entanglement
entropy'' formulated via $Z_{CFT}$ translates to a bulk object
formulated via the Wavefunction $\Psi_{dS}$ (a single ket, rather than
a density matrix), leading (not surprisingly) to non-hermitian
structures. In \cite{Narayan:2023zen}, fuelled by the analytic
continuation, a heuristic version of the Lewkowycz-Maldacena argument
\cite{Lewkowycz:2013nqa,Dong:2016hjy,Dong:2016fnf,Dong:2013qoa} (see
also \cite{Casini:2011kv}) was constructed for the no-boundary de
Sitter extremal surfaces for maximal (IR) subregions. Roughly the
boundary replica argument on $Z_{CFT}$ translates now to a bulk
replica argument on the Wavefunction $\Psi_{dS}$ which is essentially
pseudo-entropy. The area now is interpreted as the amplitude for
creation of a cosmic brane that localizes on the part Euclidean, part
timelike no-boundary extremal surface.

In this paper, we study no-boundary extremal surfaces\footnote{The
  phrase ``no-boundary extremal surface'' refers to the
  Hartle-Hawking type nature of these surfaces in no-boundary de
  Sitter space \cite{Hartle:1983ai}, not to be confused with the fact
  that the surfaces are anchored at an asymptotic (future)
  boundary. Similar comments apply to the no-boundary slow-roll
  inflation studies here.}
in slow-roll
inflation models, with certain inflationary cosmological perturbations
to no-boundary de Sitter space which preserve the spatial spherical
isometry of $dS$ in global coordinates. These perturbations are
described by scalar inflaton perturbations defined imposing regularity
at the no-boundary point: this induces corresponding metric
perturbations as well (see Figure~\ref{fig2}). The perturbations have
explicit analytic (although still adequately complicated) expressions
to $O(\epsilon)$ in the slow-roll parameter $\epsilon$, as described
in \cite{Maldacena:2024uhs} (and related previous work
\eg\ \cite{Hartle:1983ai}-\cite{Hertog:2023vot} that we found
useful). This allows us to perform explicit analytic calculations of
the no-boundary extremal surface areas and compare them with the
Wavefunction.  As in de Sitter, we consider maximal (IR) subregions on
equatorial plane slices regarded as boundary Euclidean time slices,
and look for codim-2 extremal surfaces dipping into the bulk time
direction. These are, mostly, entirely timelike in the top Lorentzian
part, going around the Euclidean hemisphere. However now the
inflationary perturbations induce small wiggles around pure de Sitter,
so the metric has various interesting real and imaginary pieces. Thus
the no-boundary extremal surface areas now have nontrivial real and
imaginary pieces which now arise from both the Euclidean hemisphere
and the Lorentzian timelike regions. In general it turns out that the
corresponding area integrals must be regarded carefully in the complex
time-plane defining appropriate contours (Figure~\ref{fig3}) that
avoid extra poles at the complexification point that arise from the
slow-roll perturbations (similar in spirit to calculations of the
semiclasical Wavefunction of the Universe).  Doing this carefully, we
eventually find divergent pure imaginary pieces from near the future
boundary as well as real and imaginary finite slow-roll corrections to
the leading half de Sitter entropy ${\pi\,l^2\over 2G_4}$ contribution
from the hemisphere.  For example, we obtain (\ref{sr4-area-final}) in
$dS_4$ slow-roll inflation with the finite parts being\
${\pi\,l^2\over 2G_4} \big(1 + \epsilon\, (\log 4 - {7\over 2}
+ i\pi)\big)$\,. These finite $O(\epsilon)$ corrections precisely
match the finite $O(\epsilon)$ corrections in the expansion of the
semiclassical Wavefunction of the Universe (equivalently the on-shell
action) in slow-roll inflation described in \cite{Maldacena:2024uhs}.
This is consistent with the Lewkowycz-Maldacena interpretation in
\cite{Narayan:2023zen} of these no-boundary extremal surface areas
giving the probability for cosmic brane creation but now in the
slow-roll no-boundary geometry.

We do a similar calculation for $dS_3$ slow-roll inflation as well,
with similar spatial spherical symmetry preserving inflaton
perturbations and corresponding metric ones. The no-boundary extremal
surface areas again have real and imaginary slow-roll corrections.
We find that the real finite $O(\epsilon)$ corrections again match
those in the expansion of semiclassical Wavefunction (or on-shell
action), but the imaginary finite parts do not match. This is likely
due to the fact that the boundary reflects a CFT on an even
dimensional sphere, with potential extra pure imaginary terms that
arise from anomalies. The probabilities, controlled by the real
parts, match.

To put this in perspective, it is worth recalling the behaviour of
minimal RT surfaces in the $AdS$ black hole
\cite{Ryu:2006bv,Ryu:2006ef}.  As the boundary subregion size
increases, the RT surface dips deeper into the bulk and in the IR
limit (maximal subregion), the surface wraps the black hole horizon.
The finite part of holographic entanglement entropy then becomes black
hole entropy which is the thermal entropy of the dual field theory:
this entropy can also be realized from the on-shell action regarded as
a partition function \cite{Gibbons:1976ue} (see also
\cite{Hawking:1982dh,Witten:1998zw}). In the de Sitter case with
surfaces anchored at the future boundary, the only turning points are
in the Euclidean hemisphere which then gives half de Sitter entropy as
the real finite part of the no-boundary extremal surface area (in
particular for maximal (IR) subregions).  This can also be realized by
evaluating the on-shell action (semiclassical Wavefunction
$\Psi_{dS}$): the real part then gives the probability $|\Psi_{dS}|^2$
controlled by de Sitter entropy. The underlying reason here in $dS$
stems from the Lewkowycz-Maldacena cosmic brane interpretation in
\cite{Narayan:2023zen}. In the context of slow-roll inflation, the
perturbations mix everything but the cosmic brane creation probability
controlled by the real finite parts of the no-boundary areas continues
to match the probability from the Wavefunction, corroborating the
replica arguments near de Sitter. Both probabilities are controlled by
de Sitter entropy and its slow-roll corrections, which is the maximum
amount of ``stuff'' in the space.

We also study Schwarzschild de Sitter black holes which in general
have conical singularities at the black hole and cosmological horizons
which have different time periodicities. In the limit of small mass
regarded as a perturbation, we define an appropriate no-boundary
geometry at the cosmological horizon whose location is shifted by the
black hole mass. Now the no-boundary extremal surface area is similar
to that in de Sitter space with a reduced cosmological scale, with the
real hemisphere contribution being half the entropy for the modified
de Sitter space.

We review de Sitter extremal surfaces in sec.~\ref{sec:dSnbRev}. The
slow-roll inflation extremal surfaces appear in sec.~\ref{sec:srnb}
and sec.~\ref{dS4srnb} for the $dS_4$ case, and sec.~\ref{dS3srnb} for
the $dS_3$ case.  We then discuss Schwarzschild de Sitter in
sec.~\ref{SdSnb} and FRW+slowroll cosmologies briefly in
sec.~\ref{FRW+SR}. Sec.~\ref{sec:Disc} contains a Discussion. Various
details appear in the Appendices, in sec.~\ref{App:dS4-sr-setup} and
sec.~\ref{App:dS3-sr-setup} on the inflation setup broadly, in
sec.~\ref{App:dS4sr} and sec.~\ref{App:dS3sr} on the area calculation,
and in sec.~\ref{App:infl-Action} on the Wavefunction/action.

\section{Reviewing de Sitter extremal surfaces}\label{sec:dSnbRev}

We briefly review \cite{Narayan:2022afv,Narayan:2023zen} and previous
work here on generalizations of RT/HRT extremal surfaces to de Sitter
space (see also \cite{Doi:2022iyj}). These involve considering the
bulk analog of setting up entanglement entropy in the dual Euclidean
$CFT$ on the future boundary \cite{Narayan:2015vda}, restricting to
some boundary Euclidean time slice, defining subregions on these, and
looking for extremal surfaces anchored at $I^+$ dipping into the time
(holographic) direction. Analysing this shows that there are no
spacelike surfaces connecting points on $I^+$. In entirely Lorentzian
$dS$, there are future-past timelike surfaces stretching between
$I^\pm$ \cite{Narayan:2017xca,Narayan:2020nsc}, akin to rotated
analogs of the Hartman-Maldacena surfaces \cite{Hartman:2013qma} in
the eternal $AdS$ black hole \cite{Maldacena:2001kr}: these have pure
imaginary area, relative to $AdS$ spacelike RT/HRT surfaces.  With a
no-boundary type Hartle-Hawking boundary condition, the top half of
these timelike surfaces joins with a spacelike part on the hemisphere
giving a complex-valued area \cite{Doi:2022iyj}, \cite{Narayan:2022afv}\
(and \cite{Hikida:2022ltr,Hikida:2021ese} for $dS_3/CFT_2$). The real
part of the area arises from the hemisphere and is precisely half de
Sitter entropy. Due to the presence of timelike components in these
extremal surfaces, these areas are best interpreted as pseudo-entropy,
as we see below. From the dual side, in various toy models of
``entanglement entropy'' in ghost-like theories, complex-valued
entropies arise naturally \cite{Narayan:2016xwq,Jatkar:2017jwz} (see
also \cite{Doi:2024nty}): the negative norm states here lead to
imaginary components. It is worth noting that the adjoints of states
are nontrivial in such ghost-like theories so these are perhaps more
correctly thought of as pseudo-entropy.
\begin{figure}[h] 
\hspace{0.5pc}
\includegraphics[width=6.5pc]{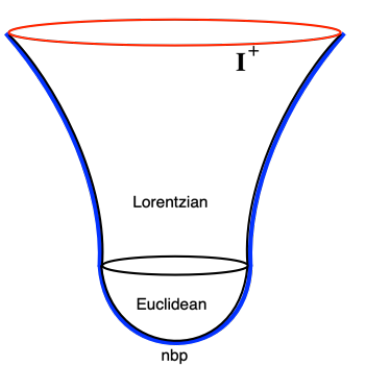}
\hspace{1pc}
\begin{minipage}[b]{29pc}
\caption{{ \label{fig1}
    \footnotesize{
      No-boundary de Sitter space, with the top Lorentzian region
      continuing smoothly into the Euclidean hemisphere region
      ending at the no-boundary point. Also shown are IR no-boundary
      extremal surfaces (blue) anchored at the future boundary $I^+$
      dipping into the time direction, timelike in the Lorentzian region
      and going around the hemisphere.
        }}}
\end{minipage}
\end{figure}

These $dS$ extremal surfaces can also be realized via analytic
continuations from $AdS$ (reviewed below). This suggests a natural way
to obtain the no-boundary de Sitter extremal surface areas through a
heuristic replica argument \cite{Narayan:2023zen} involving an
analytic continuation of the Lewkowycz-Maldacena formulation
\cite{Lewkowycz:2013nqa} in $AdS$ to derive RT entanglement entropy
(generalized in \cite{Dong:2016hjy}, \cite{Dong:2016fnf},
\cite{Dong:2013qoa}; see also \cite{Casini:2011kv} and the review
\cite{Rangamani:2016dms}). The crucial
difference here is that since the analytic continuation maps
$Z_{bulk}^{AdS}$ to the de Sitter Wavefunction $\Psi_{dS}$, this is
now a replica formulation on $\Psi_{dS}$, considering
the $dS/CFT$ dictionary $Z_{CFT}=\Psi_{dS}$\ \cite{Maldacena:2002vr}.
With the Wavefunction $\Psi_{dS}$ regarded as an amplitude (or
transition matrix from ``nothing''), this gives pseudo-entropy.
In particular the codim-2 brane that smooths out bulk (orbifold)
singularities is now a time-evolving, part Euclidean, part timelike,
brane. This gives a complex area semiclassically, with the real
part from the Euclidean hemisphere and the timelike part pure
imaginary. In this Lewkowycz-Maldacena formulation, the area of these
no-boundary $dS$ extremal surfaces arises as the amplitude for cosmic
brane creation. So it is important that the divergent pieces of the
area arising from near the future boundary are pure imaginary, since
otherwise this amplitude would diverge. As it is, there is a finite
probability: the real part arises from the maximal hemisphere, with size
set by $dS$ entropy.

Overall, directly analysing the bulk extremization and calculating the
no-boundary extremal surface areas for the IR (maximal) subregions at
the future boundary gives
\be\label{nbdS43rev}
(dS_4)\quad 
S_{nb} = -i\,{\pi l^2\over 2G_4} {R_c\over l} + {\pi l^2\over 2G_4}\,;
\qquad\quad
(dS_3)\quad 
S_{nb} = -i{l\over 2G_3}\log {R_c\over l} 
+ {\pi\,l\over 4G_3}\,,
\ee
with the pure imaginary piece from the top Lorentzian part of $dS$ and
the real piece (precisely half de Sitter entropy) from the Euclidean
hemisphere.
These can also be realized via analytic continuations from the $AdS$
RT surfaces which all lie on a constant time slice.
Under the\ $dS\leftrightarrow AdS$\ analytic continuation, this
$AdS$ constant time slice continues to $dS$ as
\bea\label{nbdSstattoAdS2}
ds^2_{(r>L)} = {dr^2\over 1+{r^2\over L^2}} + r^2 d\Omega_{d-1}^2
\ \ &\xrightarrow{\ L\ra -il\ }& \ \
ds^2_{(r>l)} = -{dr^2\over {r^2\over l^2}-1} + r^2 d\Omega_{d-1}^2 \,,\nn\\
ds^2_{(r<L)} = {dr^2\over 1+{r^2\over L^2}} + r^2 d\Omega_{d-1}^2
\ \ &\xrightarrow{\ L\ra -il\ }& \ \
ds^2_{(r<l)} =  {dr^2\over 1-{r^2\over l^2}} + r^2 d\Omega_{d-1}^2 \,.
\eea
The $AdS$ boundary at $r\ra\infty$ maps to the $dS$ future boundary
$I^+$ at $r\ra\infty$, and the $AdS$ region $r\in [L,\infty]$ maps
to the $dS$ future universe $F$ parametrized by $r\in [l,\infty]$\
(and $r$ is time here).
The $dS$ hemisphere is $\tau_E=-it=[0,{\pi\over 2}]$ where\
$-{dr^2\over {r^2\over l^2}-1}>0$\ is Euclidean.

The global $dS_{d+1}$ metric with $S^d$ cross-sections, restricted to
any equatorial $S^d$ plane is identical to the $t=const$ slice of the
(Lorentzian) $dS$ static coordinatization (\ref{nbdSstattoAdS2}),
above:
\be\label{dSglobaleqSd-dSstattconst}
ds^2_{global}\Big\vert_{\theta_d=const} = 
-d\tau^2+ l^2\cosh^2{\tau\over l}\, d\Omega_{d-1}^2\ \
=\ \ -{dr^2\over {r^2\over l^2}-1} + r^2 d\Omega_{d-1}^2 = 
ds^2_{static}\Big\vert_{t=const}\,,
\ee
using $r = l\cosh{\tau\over l}$.
Thus a generic equatorial plane in global $dS$ defines the same
boundary Euclidean time slice as the $t=const$ slice in the $dS$ static
coordinatization, and we will continue to use the parametrization
(\ref{nbdSstattoAdS2}).

The IR $AdS$ surface space spans the entire $AdS$ boundary
sphere. This continues to the IR $dS$ extremal surface (when the
subregion becomes the whole space at $I^+$), going from $r\ra\infty$
to $r=l$ as a timelike surface in the Lorentzian $dS$ region and then
in the hemisphere from $r=l$ to $r=r_*=0$ (where it turns around). The
turning point $r_*$ only exists in the Euclidean (hemisphere)
part of $dS$.

This IR surface starts at the boundary of the maximal subregion of
the $S^{d-1}$ (\ie\ hemisphere) so it is anchored on the equator of
the $S^{d-1}$ (red curve in Figure~\ref{fig1}) and wraps the
equatorial $S^{d-2}$. From (\ref{nbdSstattoAdS2}), it is clear that
the IR $dS$ extremal surface becomes a space-time rotation of that
in $AdS$.\ Its area continues as\ (with $R_c$ a cutoff at large $r$)
\bea\label{IRsurfAdSdS}
{V_{S^{d-2}}\over 4G_{d+1}} \int_0^{R_c} {r^{d-2} dr\over \sqrt{1+{r^2\over L^2}}}
\ & \xrightarrow{\, L\ra -il\,} &\
{V_{S^{d-2}}\over 4G_{d+1}} \int_0^l {r^{d-2} dr\over \sqrt{1-{r^2\over l^2}}}
+ {V_{S^{d-2}}\over 4G_{d+1}}
\int_l^{R_c} {r^{d-2} \sqrt{dr^2\over -({r^2\over l^2}-1)}} \ , \nn\\
&& =\ {1\over 2} {l^{d-1}V_{S^{d-1}}\over 4G_{d+1}} - i\# {l^{d-1}\over 4G_{d+1}}
{R_c^{d-2}\over l^{d-2}} +\, \ldots
\eea
where the $\ldots$ are subleading imaginary terms.
For $AdS_4\ra dS_4$ and $AdS_3\ra dS_3$, we obtain
\be\label{IRsurfAdSdS4}
{V_{S^1}\over 4G_4}\int_0^{R_c}{rdr\over\sqrt{1+{r^2\over L^2}}} =
{\pi L^2\over 2G_4} \Big({R_c\over L} - 1\Big)\ \ \ra\ \
-i{\pi l^2\over 2G_4} {R_c\over l} + {\pi l^2\over 2G_4} = S^{IR}_{dS_4}\,,
\ee
\be\label{IRsurfAdSdS3}
{V_{S^0}\over 4G_3}\int_0^{R_c}{dr\over\sqrt{1+{r^2\over L^2}}} =
{2L\over 4G_3}\log{R_c\over L}\ \ \ra\ \
-i{l\over 2G_3}\log{R_c\over l} + {\pi l\over 4G_3} = S^{IR}_{dS_3}\,.
\ee
These extremal surfaces can also be described explicitly in $dS_3$ for
more general boundary subregions (and for near maximal subregions in
$dS_{d+1}$). Drawing out the extremal surfaces geometrically leads to
a natural geometric identification of the bulk subregion dual to a
boundary subregion as an appropriate ``pseudo-entanglement'' wedge
defined as the appropriate domain of dependence bounded by the
boundary subregion at $I^+$ and the extremal surface. However it is
\emph{only} for maximal subregions that this leads to consistent
disjoint bulk subregions corresponding to disjoint boundary
subregions, and thereby a version of subregion-subregion duality. Thus
maximal (IR) subregions appear to be the only well-defined subregions
geometrically from the perpective of subregion duality.

\section{Slow-roll inflation, no-boundary extremal surfaces}\label{sec:srnb}

We will now describe no-boundary extremal surfaces in slow-roll
inflation, with the leading $dS_4$ area corrected to $O(\epsilon)$ in
the slow roll parameter. We first describe the basic setup, which is
essentially that described in \cite{Maldacena:2024uhs} (and related
previous work \eg\ \cite{Hartle:1983ai}-\cite{Hertog:2023vot} that
we found useful). The spacetime metric near global $dS_{d+1}$ is
\be\label{metric-atau-r}
ds^2 = -dt^2+a(t)^2 d\Omega_d^2\, =\, g_{aa} da^2 + a^2 d\Omega_d^2\,,
\ee
where we have redefined the time coordinate to be $a$ in the second
expression, which turns out to be convenient for our purposes. This
form of the metric component $g_{aa}$ applies for both the Euclidean
and Lorentzian regions of the spacetime.
For pure de Sitter in the above global coordinates, we have in
the Lorentzian region $r>1$,
\be\label{metric-atau-r2}
a(t) = l\cosh\tau \equiv l\,r\,,\qquad \tau={t\over l}\,,\qquad
g_{aa} = {1\over 1-r^2}<0\,.
\ee
The region $r<1$ gives $g_{aa}>0$ and describes the Euclidean hemisphere.
A global $S^d$ equatorial plane slice as a boundary Euclidean time
slice then resembles the $t=const$ slice in $dS$ static
coordinatization, as in (\ref{dSglobaleqSd-dSstattconst}). The
coordinate $r$ here is essentially $\al$ in \cite{Maldacena:2024uhs}. 

\begin{figure}[h] 
\hspace{0.5pc}
\includegraphics[width=10pc]{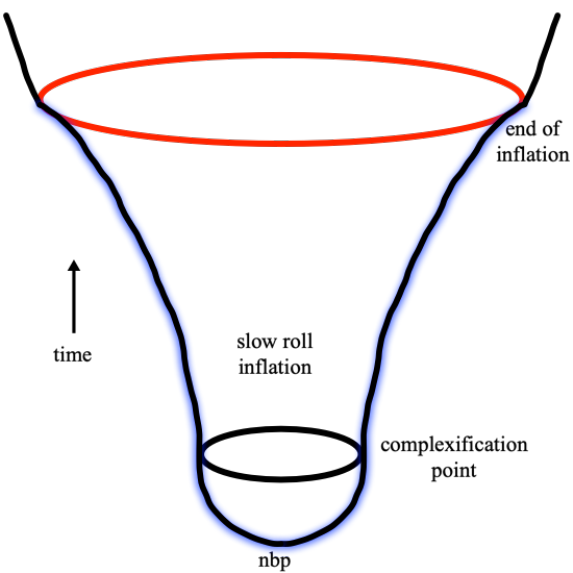}
\hspace{1.5pc}
\begin{minipage}[b]{25pc}
\caption{{ \label{fig2}
    \footnotesize{
      Slow-roll inflation as small perturbation wiggles about no-boundary
      de Sitter space. The Lorentzian region continues smoothly at the
      ``complexification point'' into the no-boundary hemisphere region
      ending at the no-boundary point.
      Inflation ends at the future boundary (the reheating surface),
      continuing with standard Big-Bang cosmology at later times. Also
      shown is the no-boundary extremal surface (blue shadow).
        }}}
\end{minipage}
\end{figure}

We now describe slow-roll inflationary perturbations about de Sitter
space (Figure~\ref{fig2}): the basic setup is in
Appendix~\ref{App:dS4-sr-setup} for the $dS_4$ case, reviewing the
description in \cite{Maldacena:2024uhs}.  The inflaton scalar field
$\phi$ rolls slowly down the potential $V(\phi)$, with a small
slow-roll parameter $\epsilon={V'^2\over 2V^2}\ll 1$.\ We use the
Einstein-scalar equations to solve for the inflaton perturbation
profile to $O(\epsilon)$ and thereby the metric correction to that
order (general useful reviews on inflation include
\cite{Baumann:2009ds,Baumann:2014nda}).

Then the metric component $g_{rr}$ to $O(\epsilon)$ in the slow-roll
correction has the form
\be\label{grr-O(epsilon)}
g_{aa} = {1\over 1-r^2}\big( 1 + 2\epsilon\,\beta_>(r) \big)\,,
\ee
where the slow roll correction $\beta_>(r)$ is a function of the
rolling inflaton profile: we will describe this in detail later.  The
no-boundary extremal surfaces we are interested in lie on boundary
Euclidean time slices taken as some $S^d$ equatorial plane
(\ie\ $S^{d-1}$) and extend out from the $S^{d-2}$ boundary of the
maximal (hemisphere) subregion on this slice, as in the pure $dS$
case: they thus wrap this $S^{d-2}$ and extend in the time direction,
making up the codim-2 surface (blue shadow in Figure~\ref{fig2}).  The
metric corrections here have complicated functional form and lead to
interesting corrections to the area as we will see.  Since the
inflationary perturbations here preserve the $S^d$ spherical symmetry,
the extremal surfaces on all such equatorial plane slices are
equivalent.

\begin{figure}[h] 
\bc
\includegraphics[width=26pc]{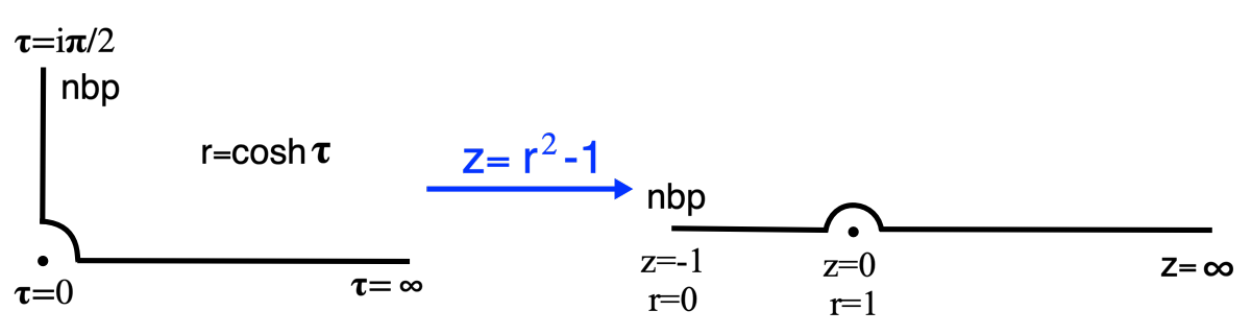}
\ec
\vspace{-5mm}
\caption{{ \label{fig3}
    \footnotesize{
      The time coordinate $\tau$ defines the contour on the left,
      from the no-boundary point in the Euclidean region at
      $\tau=i{\pi\over 2}$ to the complexification point at $\tau=0$
      and thereon to the Lorentzian region with real $\tau$ going to
      the future boundary $\tau\ra\infty$. With the coordinate
      $r=\cosh\tau$ the nbp is $r=0$ and the complexification point
      at $r=1$. In terms of the coordinate $z=r^2-1$ we have the
      nbp at $z=-1$ and the complexification point at $z=0$, giving
      the contour on the right.
        }}}
\end{figure}

The IR no-boundary extremal surface area is given by
\bea
S_{sr_{{d+1}}} &=& S_{d+1}^{r<1} + S_{d+1}^{r>1} \\
&=& {V_{S^{d-2}}\, l^{d-1}\over 4G_{d+1}} \left(
\int_0^1 {r^{d-2} \sqrt{1+2\epsilon\,\beta_<(r)}\over\sqrt{1-r^2}}\ dr\ +\
(-i) \int_1^{R_c/l}\ {r^{d-2} \sqrt{1+2\epsilon\,\beta_>(r)}\over\sqrt{r^2-1}}\ dr
\right) \nn
\eea
with the leading $dS_{d+1}$ piece and the $O(\epsilon)$ slow roll
correction, where $\beta_<(r)$ is to be obtained by analytically
continuing $\beta_>(r)$ in the Lorentzian region $r>1$ to the
hemisphere region where $r<1$.
Expanding to $O(\epsilon)$ we obtain the first slow-roll correction to
$dS_4$ and $dS_3$ as
\bea\label{slowrollAreadS4dS3}
&& S_{sr_{{4}}} \simeq\ {\pi\,l^2\over 2G_4} \left(
-i \int_1^{R_c/l} {1+\epsilon\,\beta_>(r)\over \sqrt{r^2-1}}\ r\,dr
+ \int_0^{1} {1+\epsilon\,\beta_<(r)\over \sqrt{1-r^2}}\ r\,dr \right)
\\ [1mm]
&&
S_{sr_{{3}}} \simeq\ {l\over 2G_3} \left(
-i \int_1^{R_c/l} {1+\epsilon\,\beta_>(r)\over \sqrt{r^2-1}}\ dr
+ \int_0^{1} {1+\epsilon\,\beta_<(r)\over \sqrt{1-r^2}}\ dr \right).
\label{slowrollAreadS4dS3-2}
\eea
Towards evaluating this, note that the metric (\ref{grr-O(epsilon)}) at
leading order $O(\epsilon^0)$ (\ie\ pure de Sitter) already has a pole
at $r=1$ which is the complexification point where we transit from
Lorentzian to Euclidean signature (this is $\tau=0$, using
(\ref{metric-atau-r2})).
This does not however lead to any singularities in the
extremal surface areas: from (\ref{slowrollAreadS4dS3}),
(\ref{slowrollAreadS4dS3-2}), the 
contribution near $r=1$ is\ $\int {dr\over\sqrt{\pm(1-r)}}
\sim \sqrt{1-r}$ which is nonsingular,
resulting in (\ref{IRsurfAdSdS}), (\ref{IRsurfAdSdS4}), (\ref{IRsurfAdSdS3}).
However, when this pole leads to singularities in calculations, we
should go around the pole avoiding it, \ie\ in the upper half plane:
this defines the contour in Figure~\ref{fig3} in the $\tau$- or
$r$-variables.


Considering the $O(\epsilon)$ slow-roll corrections, there are indeed
extra singular terms in the above expressions
(\ref{slowrollAreadS4dS3}), (\ref{slowrollAreadS4dS3-2}), at the
complexification point $r=1$, stemming from the extra $\beta(r)$ terms
in the integrand (see (\ref{beta>(r)dS4}), sourced by the inflaton
profile (\ref{dS4-varphitau}), using (\ref{gaa-phiV-HamC})). These
extra poles lead to divergences, similar technically to those from
poles occurring in other calculations in slow-roll inflation: see
\eg\ \cite{Maldacena:2024uhs} for some discussions on this (including
aspects of earlier numerical studies involving complex contours in the
upper half plane avoiding the pole; see \eg\ \cite{Hertog:2023vot}).
Note that since we are regarding the spacetime background here
essentially as a near $dS$ background in the slow-roll phase to
$O(\epsilon)$, the details of the full inflationary solution (and the
detailed form of the inflaton potential) are not important: thus the
pole above necessarily remains in the $O(\epsilon)$ discussions here
and must be dealt with.

In the current context, to define the extremal surface areas
correctly, we must therefore define these areas as complex integrals
in the complex time-plane, with a contour chosen to avoid the pole at
$r=1$, by going around it in the upper half plane, as in
Figure~\ref{fig3} stated above. As long as we define the complex
time-plane integral correctly, normalizing it so that the leading
expressions are consistent with the leading de Sitter areas, the
details of the contour around the pole are not important as we will
see.  In pure de Sitter, this time contour essentially encodes the
geometric picture of time and correspondingly the shape of the IR
extremal surface as in Figure~\ref{fig1}. In slow-roll, since we are
in the semiclassical regime near de Sitter, we expect that the time
contour in our $O(\epsilon)$ slow-roll case should retain this
semiclassical geometric picture of time and reflect the shape of the
slow-roll extremal surface in Figure~\ref{fig2} as small wiggles near
the de Sitter one.


Towards describing the leading $dS_{d+1}$ area as well as the $O(\epsilon)$
slow-roll correction, we will find it convenient to use the variable
$z=r^2-1$, although our analysis can be carried out in the $r$-variable
as well (as we will describe later). The no-boundary point and the
complexification point in the $r$- and $z$-variables are
(Figure~\ref{fig3})
\be\label{z-r}
z=r^2-1\quad \Rightarrow\quad ({\rm nbp})\ \ r=0\ \equiv\ z=-1\,;\quad
({\rm complexification})\ \ r=1\ \equiv\ z=0\,.
\ee
The area integral is then recast as a complex contour integral: 
the contour $C$ in the $z$-variable, with the pole $z=0$ regulated to
$z=\delta$, is
\be
C\ =\ \left[-1,-\delta\right]\ \ \cup\ \
\left[ z=\delta e^{i\theta}\,; \theta=\pi,\ \theta=0 \right]\ \ \cup\ \
\left[\delta, z_c\right]\ .
\ee
The middle leg is a semicircle of radius $\delta$ going around the
pole, defined as going from $\theta=\pi$ to $\theta=0$\,: then
$z=\delta\,e^{i\pi}=-\delta$ at the left end of the semicircle (on
the side of the nbp) and $z=\delta$ at its right end, around the
complexification point. The details of this regulating semicircle
skirting the pole at $r=1$ or $z=0$ are not important. In essence, we
are evaluating two separate area contributions, from the Lorentzian
and hemisphere regions, with regularization at the complexification
point.

\subsection{$dS_4$ area including $O(\epsilon)$ correction}\label{dS4srnb}

We will now study the no-boundary extremal surface area
(\ref{slowrollAreadS4dS3}) in 4-dim slow roll inflation: the basic
setup here appears in App.~\ref{App:dS4-sr-setup}.  The metric is of
the form (\ref{metric-atau-r}), (\ref{metric-atau-r2}), with $d=3$,
and the overall picture is as depicted in Figure~\ref{fig2}.

The slow-roll equations of motion (\ref{dS4eoms-sr}) in the slow-roll
approximation become (\ref{dS4eoms-sr-i}), \ie\ 
\be
3 H^2 \sim V(\phi), \qquad 3 H\dot{\phi} \sim - V'(\phi)\,.
\ee
The metric component $g_{aa}$ can be shown to take the form (see
App.~\ref{App:infl-Action})
\be\label{gaa-phiV-HamC}
g_{aa} = {3-{1\over 2} (a\,\del_a\phi)^2\over 3-V(\phi)\,a^2}\,.
\ee
Expanding to $O(\epsilon)$ in the slow-roll correction with\
$\phi=\phi_*+\varphi(\tau)$\ and\ $V(\phi)=V_*+{V_*}^\prime \varphi(\tau)$,
we solve the inflaton equation of motion (\ref{sr-dS4-eom}) for the
perturbation, imposing regularity at
the no-boundary point and matching with the late time expansion
(as described in \cite{Maldacena:2024uhs}). 
This finally gives the 4-dim slow-roll inflaton profile in the $r>1$
Lorentzian region as
\be\label{dS4-varphitau}
\epsilon\equiv \frac{{{V_*}^\prime}^2}{2V_*^2}\,;\qquad
\varphi(r) = \sqrt{2\epsilon}\,\tilde{\varphi}(r)\,,\qquad
\tilde{\varphi}(r) =  \frac{1+i \sqrt{r^2-1}}{r^2}-\log \left(1-i \sqrt{r^2-1}\right)-\frac{i\pi}{2}\,,  
\ee
and the metric component $g_{rr}=g_{aa}l^2$ in the form (\ref{grr-O(epsilon)}),
where in this 4-dim case we have
\be\label{beta>(r)dS4-0}
\beta(r) = -\frac{1}{6} \left(r \frac{\partial \tilde{\varphi}}{\partial r}\right)^2
 + \frac{\tilde{\varphi}\, r^2}{1-r^2}\,.
\ee
Inputting the inflaton profile (\ref{soln-sr-dS4-eom-i}) with $r=\cosh\tau$
gives $\beta(r)$ in the $r>1$ Lorentzian region,
{\small
\be\label{beta>(r)dS4}
\beta_>(r) =
\frac{8-9 r^4+4 i r^2 \sqrt{r^2-1}+8 i \sqrt{r^2-1}-6 i r^4 \sqrt{r^2-1}+r^6 \big(6 \log \big(1-i \sqrt{r^2-1}\big)-1+3 i \pi \big)}{6 r^4 \left(r^2-1\right)}\,.
\ee }
To continue this to the $r<1$ hemisphere region, we note that it is
adequate to replace $-i\sqrt{r^2-1}$ by $\sqrt{1-r^2}$ in $\beta_>(r)$:
this defines $\beta_<(r)$.

The above recasting gives the $dS_4$ slow-roll inflation area to
$O(\epsilon)$ as
\be\label{sr4-area}
S =
{\pi\,l^2\over 2G_4} \int_C {1 + \epsilon\,\beta_>(r)\over 2\sqrt{1-r^2}}\
r\,dr\ \equiv\ 
{\pi\,l^2\over 2G_4} \int_C {1 + \epsilon\,\beta_>(z)\over 2\sqrt{-z}}\ dz\,.
\ee
For the leading pure de Sitter piece, there are no singularities
and the areas can be evaluated as separate integrals in the
Lorentzian and Euclidean regions, as in (\ref{IRsurfAdSdS}).
However for the $O(\epsilon)$ contribution to the area, it turns out
that there are extra pole pieces at the complexification point $r=1$
($z=0$). Thus it is important to define these as a contour integral
in the complex time plane, defining a contour to avoid the pole
at $r=1$, as shown in Figure~\ref{fig3}. We describe the detailed
calculation in Appendix~\ref{App:dS4sr}, in both the $z$- and
$r$-coordinates.

Adding up all the contributions along the full contour $C$ keeping
the leading order and $O(\epsilon)$ pieces in (\ref{sr4-area}), we
obtain
\bea\label{sr4-area-1}
S &=& {\pi\,l^2\over 2G_4} \Bigg[ 1-\sqrt{\delta}\ +\ 
  \epsilon \left[ \left( \log 4 - {7\over 6} + i\pi \right) -
    \left( -{2-3i\pi\over 6 \sqrt{\delta}} + {5\over 3}\right)\right]\nn\\
&& \quad +\ \sqrt{\delta} (-i+1)\ +\ \epsilon
\left[ {2-3i\pi\over 6i\sqrt{\delta}}
  - {2-3i\pi\over 6\sqrt{\delta}}\right]\\
&& \quad +\ {1\over i} (\sqrt{z_c} - \sqrt{\delta})\
+\ \epsilon \left[  \left( 1 + i {7\over 6} \sqrt{z_c}
  -i\sqrt{z_c}\log\sqrt{z_c} \right)
  - \left( {2-3i\pi\over 6i \sqrt{\delta}} + {5\over 3} \right) \right] \nn
\Bigg]\,.
\eea
Finally, note that although we have expressed the area as a single
complex-time-plane integral, we have evaluated it as separate 
integrals in the top timelike part and the
hemisphere part, with the semicircle contour only serving to avoid
the $r=1$ ($z=0$) pole.
Note that all the singular pieces near the pole cancel as they
should, so the details of the contour around the pole are unimportant.
We obtain finally 
\be\label{sr4-area-final}
S_{sr_4} = {\pi\,l^2\over 2G_4} \left( -i{R_c\over l} + 1\right)\
+\ \epsilon\, {\pi\,l^2\over 2G_4} \left( -i{R_c\over l} \log {R_c\over l}\
+ i {7\over 6} {R_c\over l}\ +\ \log 4 - {7\over 2} + i\pi  \right) ,
\ee
to $O(\epsilon)$ in the slow-roll corrections. Note that the various
finite pieces conspire to give the finite value above: the real part
of this finite value is negative.

Unlike in the pure de Sitter case, where there was a clean separation
between the real part of the area arising from the hemisphere and the
imaginary part from the top timelike Lorentzian region, the slow-roll
corrections contain real and imaginary parts from both the timelike
and hemisphere regions. The finite terms in $S_{sr_4}^\epsilon$ in
particular arise from the entire surface, both timelike and
hemisphere parts.

Recalling the comments after (\ref{slowrollAreadS4dS3}),
(\ref{slowrollAreadS4dS3-2}), the minimal time contour
(Figure~\ref{fig3}) we have employed is natural given that the
slow-roll case is near de Sitter: it faithfully dovetails with the
geometric picture of time in the semiclassical near $dS$ regime here,
and encodes the shape (Figure~\ref{fig2}) of the slow-roll IR extremal
surface which is expected to be near that in pure $dS$
(Figure~\ref{fig1}). Dramatically changing the contour would deviate
substantially from the semiclassical geometry here (also the Cauchy
theorem cannot apply given the non-analytic expressions here).

It is now interesting to note that this finite value above matches
precisely with the finite part in (\ref{I-final-i}) of the $O(\epsilon)$
slow-roll correction in the Wavefunction of the Universe $\Psi\sim e^{iI}$
in slow-roll inflation, given in \cite{Maldacena:2024uhs}: we review
the details of this calculation in App.~\ref{App:infl-Action}. But to
summarize, the on-shell $dS_4$ action with $O(\epsilon)$ slow-roll
correction is
\bea\label{sr4-action-f}
&& iI_{sr4} = {\pi\,l^2\over 2G_4} \Biggl[ 1
  +\epsilon\left(\log 4-\frac{7}{2}+i\pi\right) \nn\\
&& \qquad\qquad\qquad\qquad   - i\Big( r_c^{3}-\frac{3}{2}r_c \Big) +
  i\epsilon \left(
    r_c^3\Big(\log r_c-\frac{1}{6}\Big)
    +\frac{r_c}{4}\left(6\log r_c-11\right)
  \right) \Biggr]\,,\quad
\eea
where $r_c\equiv {R_c\over l}$ is the cutoff near the future boundary.
The first line is mostly real while the second line is pure imaginary.
We note that $|\Psi|^2\sim e^{{\pi l^2\over G_4} (1+\epsilon\,\#)}$ is controlled
by the the real terms in the first line and is the probability for
creation of this universe with slow-roll corrections. The imaginary
terms cannot agree since the area corresponds to a codim-2 extremal
surface with corresponding divergences while the action corresponds to
the full 4-dim space with its corresponding divergences near the
future boundary.

The fact that the real terms in $S_{sr_4}$ in (\ref{sr4-area-final})
match those in $iI_{sr4}$ in (\ref{sr4-action-f}) is consistent with
and corroborates the heuristic Lewkowycz-Maldacena interpretation in
\cite{Narayan:2023zen} of these no-boundary extremal surface areas as
the amplitude for codim-2 cosmic brane creation (reviewed briefly after
Figure~\ref{fig1}). More specifically
the probability for this maximal cosmic brane creation matches that
for the creation of the Universe. This probability must
be finite since it is ultimately dictated by the upper bound set by
de Sitter entropy (the maximal amount of ``stuff'' in this space,
controlled by the no-boundary maximal hemisphere): thus the divergent
terms near the future boundary cannot contribute and must be pure
imaginary. With the slow-roll corrections included, we have seen
that the finite terms arise from everywhere. Note that since
$\log 4-{7\over 2}<0$, the no-boundary area or the cosmic brane
creation probability has decreased. This is perhaps a reflection of
the fact that excitations in de Sitter space decrease its entropy (as
for a black hole in de Sitter space).

It is interesting to ask what the analog in $AdS$ is of this matching
of the real finite part of the areas and the entropy contribution in the
Wavefunction.
The $AdS$ Ryu-Takayanagi formulation relies on the existence of a nice
optimization: minimal surfaces start at the boundary and turn around
in the interior. With regard to the current context, we recall that
the RT surface on a constant time slice in the $AdS$ black hole/brane
wraps the horizon as the boundary subregion becomes the whole space,
\ie\ the maximal (IR) subregion. The finite part of the area then
becomes the bulk entropy (which for the $AdS$ black brane is the
thermal entropy in the dual field theory). This entropy can also be
realized by the well-known Gibbons-Hawking procedure of regarding the
on-shell action as a partition function \cite{Gibbons:1976ue} and
evaluating the corresponding entropy: it usefully captures phenomena
in $AdS/CFT$ such as the Hawking-Page transition
\cite{Hawking:1982dh,Witten:1998zw}. In the de Sitter case,
there are no turning points in the Lorentzian region as we have
reviewed: no-boundary extremal surfaces turn around only in the Euclidean
hemisphere and give half $dS$ entropy as the real finite part. This
entropy can also be realized from the on-shell action regarded as
a partition function: with the slow-roll corrections we obtain the
above. The fact that the maximal cosmic brane probability matches
the probability from the Wavefunction then corroborates the 
Lewkowycz-Maldacena arguments in \cite{Narayan:2023zen}.

It is worth noting that the divergent terms in the area
(\ref{sr4-area-final}) in the slow-roll corrections exhibit a
logarithmic enhancement scaling as $R_c\log R_c$, which might seem
surprising considering that the metric component $g_{aa}$ itself
(\ref{grr-O(epsilon)}) has only subleading pieces near the future
boundary. However we note that the slow-roll correction $\beta_>(r)$
(\ref{beta>(r)dS4}) part in $g_{aa}$ does have a term scaling as
$\log r$ for large $r$: thus the Lorentzian part of the area
(\ref{slowrollAreadS4dS3}) of this codim-2 surface (which has extra
$r$-factors) for large $r$ scales as $\epsilon\log r$ which leads to
the log-enhancement. Similar features appear also in the action
(\ref{sr4-action-f}) where the slow-roll correction scales as
$\epsilon r_c^3\log r_c$ while the leading term scales as $V_{S^3}\sim
r_c^3$ (as noted in \cite{Maldacena:2024uhs}).\\
(A related point is whether the logarithmic term renders the finite
piece ambiguous: in this regard note that rescaling the cutoff as
$R_c\ra \al R_c$ in (\ref{sr4-area-final}) gives an extra\
$\epsilon {\pi l^2\over 2G_4} (-i{R_c\over l}\al\log\al)$\
contribution from the $R_c\log R_c$ term, which modifies only the
leading area law divergence term. There is no standalone $\log R_c$
term, so the finite piece is unambiguous.)

To put this enhanced logarithmic scaling in perspective, it is worth
recalling similar features in $AdS/CFT$, especially in the context of
nonrelativistic condensed matter-like generalizations and Fermi
surfaces. In particular it was noted that certain classes of
hyperscaling violating Lifshitz theories exhibit novel holographic
entanglement scaling \cite{Ogawa:2011bz,Huijse:2011ef,Dong:2012se}.
While these are phenomenological gravitational descriptions, certain
gauge-string realizations were found in \cite{Narayan:2012hk}
which precisely recover this logarithmic scaling. 
For instance, the $AdS_5$ plane wave\ 
$ds^2 = {R^2\over r^2} [-2dx^+dx^- + dx_i^2 + dr^2] + R^2Qr^{2} (dx^+)^2
+ R^2d\Omega_5^2$ is interpreted as a simple anisotropic state in
the dual ${\cal N}=4$ Super Yang-Mills CFT$_4$ with holographic
energy momentum density $T_{++}\sim Q$\,. Under $x^+$-dimensional
reduction this gives\ $ds^2=r^{2 \theta \over d_i} \big(-{dt^2 \over r^{2z}} + 
{\sum_{i=1}^{d_i}  dx_i^2 + dr^2 \over r^2 }\big)$\ with\
$d_i=2,\ \theta=1,\ z=3$ (and $x^-\equiv t$). Holographic
entanglement entropy \cite{Narayan:2012ks} (see also
\cite{Narayan:2014ofl}) for a strip subregion along the flux
direction then gives the
leading divergence\ $S^{div}\sim N^2{V_2\over\epsilon^2}$\,,\ and
the finite part\ $S^{fin}\sim N^2V_2\sqrt{Q}\log (lQ^{1/4})$\,.
Identifying the energy flux $Q$ and the UV cutoff $\epsilon$ with
the Fermi momentum\ $k_F\equiv Q^{1/4}\sim {1\over\epsilon}$\
recasts the area law divergence as\ $S^{div}\sim N^2V_2k_F^2$\ with
the finite part resembling the logarithmic enhancement for a Fermi
surface\ $S^{fin}\sim N^2V_2k_F^2\log (l k_F)$\,.
Although the $AdS/CMT$ context is quite different, we see that
there are structural similarities with the extremal surface areas
(\ref{sr4-area-final}). These complex areas here are expected to encode
boundary entanglement entropy in the exotic Euclidean (ghost-like)
CFT$_3$ dual to $dS_4$ under the slow-roll deformations sourced by the
inflaton (although understanding this nontrivial dual in
detail is challenging!).

Finally one might ask what the no-boundary slow-roll
corrected areas map to in $AdS$ under the analytic continuation $L\ra
-il$ which for the maximal subregions in $AdS\leftrightarrow dS$ are
given by (\ref{IRsurfAdSdS}), (\ref{IRsurfAdSdS4}),
(\ref{IRsurfAdSdS3}). Then the leading term in the area
(\ref{sr4-area-final}) is simply the $AdS$ area in (\ref{IRsurfAdSdS4}),
but the analytic continuation of the slow-roll correction gives
imaginary components as well. This suggests that the corresponding
perturbations in $AdS$ induce timelike components to the extremal
surface. However this blithe continuation needs to be examined
more carefully requiring physically reasonable scalar perturbations
in $AdS$ that might be analogous to the inflationary no-boundary de
Sitter perturbations here.

\section{Other cosmologies}

We will now describe no-boundary extremal surfaces in some other
cosmological spacetime backgrounds, including in particular $dS_3$
slow-roll inflation, and Schwarzschild de Sitter black holes with
small mass.

\subsection{$dS_3$ area including $O(\epsilon)$ correction}\label{dS3srnb}

We can generalize the above 4-dim analysis in sec.~\ref{dS4srnb} to
3-dimensions and study 3-dim slow roll inflation: the basic setup here
appears in App.~\ref{App:dS3-sr-setup}. The 3-dim gravity case has no
intrinsic dynamics but the inflaton scalar degree of freedom makes it
more interesting, even if this is just a toy model of slow-roll
inflation.  The metric is of the form (\ref{metric-atau-r}),
(\ref{metric-atau-r2}), with $d=2$, and the overall picture is as
depicted in Figure~\ref{fig2}.

Here we have the slow-roll equations of motion (\ref{dS3eoms-sr}),
which in the slow-roll approximation become (\ref{dS3eoms-sr-i}), \ie\ 
\be
H^2 \sim V(\phi), \qquad 2 H\dot{\phi} \sim - V'(\phi)\,.
\ee
The inflaton equation of motion (\ref{sr-dS3-eom}) for the perturbation
can be solved as in the $dS_4$ slow-roll case, imposing regularity at
the no-boundary point and matching with the late time expansion. 
This finally gives the solution as (\ref{soln-sr-dS3-eom-ii}), \ie\
\be
\varphi(\tau)= \sqrt{2\epsilon}\Biggl[\left(\frac{i\pi}{4}-\frac{\tau}{2}\right) \tanh\tau+\frac{\log2}{2}-\frac{i\pi}{4}\Biggr] \equiv
\sqrt{2\epsilon}\, {\tilde\varphi(\tau)}\,.
\ee
The slow-roll correction to the metric here, using $r=\cosh\tau$, is
\be
g_{rr} = {1 - {1\over 2} (r\del_r\varphi)^2\over 1- V r^2}
\simeq {1\over 1-r^2} \left( 1 + 2\epsilon \beta_>(r) \right)
\ee
where
\be
2\epsilon \beta_>(r) \equiv {{V'_*}^2\over V_*^2}
\left( -{1\over 2}(r\del_r{\tilde\varphi})^2 + {{\tilde\varphi}\ r^2\over 1-r^2} \right)
\ee
evaluates to the expression in (\ref{beta>(r)dS3}).

The slow-roll $dS_3$ inflation no-boundary extremal surface area
in (\ref{slowrollAreadS4dS3-2}) is
\be
S_{sr_{{3}}} \simeq\ {l\over 2G_3} \left(
\int_\delta^{z_c} {1+\epsilon\,\beta_>(z)\over 2i \sqrt{z (1+z)}}\ dz
+ \int_\delta^{1} {1+\epsilon\,\beta_<(z)\over 2 \sqrt{z (1-z)}}\ dz \right)
+ {l\over 2G_3} I_\epsilon^\theta
\ee
where $I_\epsilon^\theta$ is the contribution from the semicircle part
of the contour, skirting around the pole at $z=0$. We have written
this in terms of the variable $z=r^2-1$ in (\ref{z-r}) for $z>0$
and then continued as $z\ra -z$ to the hemisphere region, as in the
$dS_4$ case.

We describe the details of this calculation in Appendix~\ref{App:dS3sr}.
From there, we obtain
\bea\label{sr3-area-1}
S\! &=&\! {l\over 2G_3} \Bigg[ {\pi\over 2}  
+ \epsilon \Big[ \frac{1}{16} i \left(-(\log z_c)^2+\log z_c+\pi ^2 
    + 4 i \pi + 4 (\log 2)^2-6 \log 2 \right) \nn\\
&& \qquad\qquad\qquad
    - \left(-\frac{\pi }{8} + \frac{i \left(8 i \pi +\pi ^2
      -4 \log (16)\right)}{32 \sqrt{\delta}} \right) \Big]   \Bigg]\
+\ I_\epsilon^\theta  \\
&& 
-\, i \log\sqrt{z_c} 
+ \epsilon \Big[ {\pi\over 16} (2\pi i - 1 -\log 16)
  - \left( -\frac{\pi }{8}+\frac{ \left(8 i \pi +\pi ^2
  -4 \log (16)\right)}{32 \sqrt{\delta}} \right) \Big] \Bigg] \nn
\eea
which simplifies to (after rewriting using $z=r^2-1$ and expanding at
large $r_c={R_c\over l}$)
\bea\label{sr3-area-final}
S \!&=&\! {l\over 2G_3}
\Big( {\pi\over 2} -i\log{R_c\over l} \Big) \\
&& +\ \epsilon\,{l\over 2G_3}
\left( - {\pi\over 16} (1 + \log 16)\ +
\ \frac{i}{16} \big(2 \log {R_c\over l}-4\Big(\log {R_c\over l}\Big)^2+3\pi ^2 
  + 4 (\log 2)^2-6 \log 2 \big) \right) . \nn
\eea
The real part of the area including the slow-roll correction here
is\
\be\label{sr3-area-final-real}
{\rm Re} S =
{\pi\,l\over 4G_3} - \epsilon\,{l\over 2G_3} {\pi\over 16} (1 + \log 16)\,.
\ee
As in $dS_4$ inflation, this in fact matches the real parts in
(\ref{srdS3action0}) and (\ref{srdS3action-epsilon}) in
App.~\ref{App:infl-Action} obtained to $O(\epsilon)$ from the
semiclassical expansion of action or the the logarithm of the
Wavefunction, \ie\ $\log\Psi\sim iI_{cl}$\,.  Thus the maximal
cosmic brane creation probability matches the probability $|\Psi|^2$
for Universe creation, corroborating the heuristic Lewkowycz-Maldacena
interpretation in \cite{Narayan:2023zen} of these no-boundary extremal
surface areas.

The details of this calculation appear in App.~\ref{App:infl-Action}.
The imaginary finite pieces in (\ref{srdS3action0}),
(\ref{srdS3action-epsilon}) do not agree. The dual here is an
even-dimensional CFT and anomaly contributions of the CFT on a
2-sphere are expected to give further pure imaginary contributions.
The CFT anomaly is expected to be proportional to the central charge
which in $dS_3/CFT_2$ is pure imaginary. It would be interesting to
understand this breaking of dual conformal invariance by such
inflationary perturbations in greater detail.

In this light, it is worth looking more closely at the leading term
as well. This in fact contains a subleading finite term near the
future boundary: we have the timelike part of the area\
$- i {l\over 2G_3} \int_1^{R_c/l} {dr\over\sqrt{r^2-1}}
= -i {l\over 2G_3} \log {R_c\over l} - i {l\over 2G_3} \log 2$\,,
noting that the indefinite integral is $\log (r+\sqrt{r^2-1})$.
This subleading finite term is pure imaginary so the real part of
the area controlling the cosmic brane creation probability remains
half $dS_3$ entropy.

\subsection{Schwarzschild de Sitter black holes with small mass}\label{SdSnb}

We will discuss Schwarzschild de Sitter black holes here, but with the
black hole mass treated as perturbation: some useful references include
\cite{Bousso:1995cc,Bousso:1996au}, as well as our previous work
\cite{Fernandes:2019ige,Goswami:2022ylc,Goswami:2023ovb}.
The metric is
\be\label{SdS4metric}
ds^2 = -f(r) dt^2 + {dr^2\over f(r)} + r^2 d\Omega_2^2\,,\qquad
f(r) = 1-{2m\over r}- {r^2\over l^2}\,.
\ee
The metric function can be written as
\be\label{f(r)SdS4roots}
f(r) = {1\over l^2\,r} (r_D-r)(r_D+r_S+r)(r-r_S)\,,\qquad
r_Dr_s(r_D+r_S)=2ml^2\,,\quad r_D^2+r_Dr_S+r_S^2=l^2\,.
\ee
The roots $r_S$ and $r_D$ correspond to the black hole and cosmological
horizons respectively, and satisfy the constraint relations above,
alongwith $0\leq r_S\leq r_D\leq l$\,. In addition, we have the
condition ${m\over l}\leq {1\over 3\sqrt{3}}$ for a physical black
hole horizon to exist.

In the limit of small mass $m\ll l$, the roots representing the horizon
locations can be approximated as
\be
r_D\sim l-G_4m\,,\qquad r_S \sim 2G_4m\,,
\ee
so that the cosmological horizon at $r_D=l$ shifts a little inward
for a nonzero black hole horizon.  Since there is a curvature singularity at
$r=0$ inside the black hole horizon, strictly speaking there is no
smooth Euclidean instanton that can be extended to a smooth
no-boundary geometry. However treating the black hole perturbation to
only serve to shift the cosmological horizon, we can look for a
no-boundary geometry continuing to take the no-boundary point as
$r=0$. In other words, we disregard the spacetime geometry effects
of the black hole (and thus disregard the singularity etc). This
amounts to approximating the metric function as
\be\label{SdS4-f(r)approx}
f(r) = {1\over l^2} \left(r_D^2 - r^2 + r_S(r_D-r) \right)
\left( 1 - {r_S\over r}\right)\ \
\xrightarrow{\,r_S\ra 0,\ \ \ l\sim r_D\,}\ \ 1 - {r^2\over r_D^2}\,.
\ee
Now we Euclideanize $t=i\tau_E$, and for $r<r_D$ we glue on a
Euclidean hemisphere which is smooth at the no-boundary point $r=0$,
and thus replaces the earlier $SdS$ geometry.

On the $t=const$ slice taken as a boundary Euclidean time slice, we
consider the maximal (hemisphere) subregion on the $S^2$ at the future
boundary.  Then the no-boundary area becomes
\be
S = {V_{S^1}\over 4G_4}
\int_0^{r_D} {r\,dr\over \sqrt{1 - {r^2\over r_D^2}}}\ +\
{(-i)V_{S^1}\over 4G_4}
\int_{r_D}^{R_c} {r\,dr\over \sqrt{{r^2\over r_D^2} - 1}}\,.
\ee
This gives
\be
S =  - i{\pi\,r_D^2\over 2G_4}\,{R_c\over r_D} + {\pi\,r_D^2\over 2G_4}\,.
\ee
The real part of this area is half the entropy of the shifted
cosmological horizon with size $r_D\sim l-G_4m$\,: in other words,
treating this as just a modified no-boundary de Sitter space, we
obtain the earlier result for pure $dS_4$.
To $O(m)$, this gives
\be
{\rm Re} S \sim {\pi\over 2G_4} (l-G_4m)^2\ \sim\ {\pi\,l^2\over 2G_4}
- \pi\,l\,m\,,
\ee
This matches half the Euclidean action, expanded to $O(m)$, for
Schwarzschild de Sitter black holes obtained in \cite{Morvan:2022ybp}
by a careful treatment of the conical singularities at the two
horizons incorporating $\delta$-function sources there.
However it is worth emphasizing that the extremal surface area has
been obtained under the approximation (\ref{SdS4-f(r)approx}). It
would be worth refining this more carefully.

For 3-dim Schwarzschild de Sitter given by
\be
ds^2 = -f(r) dt^2 + {dr^2\over f(r)} + r^2 d\phi^2\,,\qquad
f(r)=1-8G_3E-{r^2\over l^2}\,.
\ee
In this case, there is only one horizon at $r_D=l\sqrt{1-8G_3E}$\,.
At sufficiently large $E$, the space closes up and the entropy
${\pi\,r_D\over 2G_3}$ vanishes.
As before, for small energy $E$, we approximate the metric
function as\ $f(r) \sim 1-{r^2\over r_D^2}$ and evaluate the
hemisphere contribution to the no-boundary area as
\be
S^h = {2\over 4G_3} \int_0^{r_D}
{dr\over\sqrt{1-{r^2\over r_D^2}}} = {\pi\,r_D\over 4G_3}\,,
\ee
which is half the entropy of the modified $dS_3$ space (compare
(\ref{nbdS43rev})).
However it is worth noting that the area calculation done exactly
instead gives
\be
S'^h = {2\over 4G_3} \int_0^{r_D}
{dr\over\sqrt{1-8G_3E-{r^2\over l^2}}} = {\pi\,l\over 4G_3}\,,
\ee
which is half $dS_3$ entropy independent of the mass $E$. It is
important to note however that in this case, the location $r=0$
is a conical singularity as can be seen from the metric there, \ie\
$ds^2 \sim -{r_D^2\over l^2} dt^2 + {l^2 dr^2\over r_D^2} + r^2d\phi^2
\sim -dt'^2 + dr'^2 + r'^2d\phi'^2$\ redefining $t'={r_D\over l}t\,,\
r'= {lr\over r_D}\,,\ \phi'={r_D\over l}\phi$\ (see \eg\
\cite{Spradlin:2001pw}). Thus we see that for $r_D<l$, the angle
$\phi'$ is identified modulo $2\pi {r_D\over l}$ reflecting the
conical singularity at $r=0$. Thus the area $S'^h$ pertains to this
singular space while $S^h$ pertains to the no-boundary space obtained
from $SdS_3$ under the approximations of $f(r)$.

Thus overall, it is important to note that Schwarzschild de Sitter
spaces for generic mass have distinct Euclidean periodicities at the
black hole and cosmological horizons so they cannot be made regular
except at the extremal Nariai point. Thus in evaluating these
no-boundary extremal surface areas we are making approximations
to construct appropriate no-boundary geometries for small mass
perturbations near de Sitter.
Perhaps one may better define these areas by more careful treatment
of the conical singularities for generic mass.

\subsubsection{Extremal $SdS_4$: Nariai}

The Nariai or extremal limit of Schwarzschild de Sitter gives a
near horizon $dS_2\times S^2$ geometry. This occurs when $r_D=r_S$:
then (\ref{f(r)SdS4roots}) gives $r_D=r_S=r_0={l\over\sqrt{3}}$ and
${m\over l}={1\over 3\sqrt{3}}$\,. This is a degenerate limit but
regulating near Nariai as $r_D=r_0-x,\ r_S=r_0+x$, gives
\be
f(r) \sim {3\over l^2} (x^2-(r-r_0)^2)\,,
\ee
where we approximate $r\sim r_0$ in the $(r_D+r_S+r)$ factor. This
amounts to zooming in to the extremal throat region, with the
transverse $S^2$ of size $r_0={l\over\sqrt{3}}$.
Defining
\be
\psi = \int {dr\over \sqrt{f(r)}} = {l\over\sqrt{3}} \sin^{-1}{r-r_0\over x}
\equiv {l\over\sqrt{3}} {\tilde\psi}\,,
\ee
the metric on the $t=const$ slice approximates to
\be
ds^2 \sim d\psi^2 + {l^2\over 3}d\Omega_2^2\,.
\ee
For the no-boundary hemisphere we have
${\tilde\psi}\in [0,{\small {\pi\over 2}}]$, so the area for a
hemispherical cap on an equatorial plane in the $S^2$ is approximated as
\be
S \sim {l^2\over 3} \int_0^{{\pi\over 2}} d{\tilde\psi}\
{V_{S^1}\over 4G_4} =  {\pi\over 6}\,{\pi\,l^2\over 2G_4}\,,
\ee
which is a factor ${\pi\over 6}\sim {1\over 2}$ times $dS_4$ entropy.
On the other hand, the Euclidean action for the Nariai limit
$dS_2\times S^2$ is a factor ${2\over 3}$ times that for $dS_4$
\cite{Bousso:1995cc,Bousso:1996au}. This mismatch is perhaps not
surprising since the Nariai limit is far from pure de Sitter,
unlike the previous cases we have discussed.

\subsection{FRW+slow-roll cosmologies}\label{FRW+SR}

Generic FRW-type cosmologies with spatial $S^d$ slices are of the form
\be\label{FRWmet}
ds^2 = -dt^2 + a^2(t) d\Omega_d^2\,,\qquad a(t) \sim t^\nu\,.
\ee
From the point of view of the slow-roll inflation phase evolving to an
FRW cosmology epoch (schematically as in Figure~\ref{fig2}), we are
interested only in regarding this as a Lorentzian FRW space with the
timelike part of the slow-roll inflation extremal surfaces joining at
the reheating surface (approximately the future boundary) with timelike
extremal surfaces in the FRW phase.  We restrict to a boundary
Euclidean time slice as some $S^d$ equatorial plane and pick maximal
(hemisphere) subregions.  Then the Lorentzian FRW region gives
timelike extremal surfaces with area (suppressing intrinsic FRW
lengthscales)
\be
S_{FRW} = -i {V_{S^{d-2}}\over 4G_{d+1}} \int_{t_0}^{t_l} dt\ a(t)^{d-2}\
\xrightarrow{\,4-dim\,}\
-i{\pi\over 2G_4} \big( t_l^{\nu+1} - t_0^{\nu+1} \big)\,.
\ee
Appending (\ref{sr4-area-final}), the full extremal surface area for
the FRW+inflation cosmology becomes $S_{FRW}+S_{sr4}$.
It may be interesting to understand if these sorts of codim-2 area
observables can be usefully employed to understand aspects of
cosmological observations.

Considering the FRW spaces (\ref{FRWmet}) in themselves, in general
these will not admit any smooth HH no-boundary Euclidean
continuation. Blithely calculating the IR no-boundary extremal surface
area gives the Euclidean hemisphere contribution as
\bea
&& S = {V_{S^{d-2}}\over 4G_{d+1}} \int_0^{\tau_l} d\tau_E\ a(\tau_E)^{d-2} 
= {V_{S^{d-2}}\over 4G_{d+1}} \int_0^{\tau_l}
{da \ a(\tau_E)^{d-2}\over (da/d\tau_E)} \nn\\
&& \quad \sim {V_{S^{d-2}}\over 4G_{d+1}} \int_0^{\tau_l} d\tau_E\, \tau_E^{(d-2)\nu}
\xrightarrow{4-dim}\ \ {\pi\over 2G_4}\, {\tau_l^{\nu+1}\over \nu+1}\,.
\eea
where we have taken $a(\tau_E)\sim \tau_E^\nu$,
with $\nu>0$. The sphere $S^d$ shrinks to zero size at the no-boundary
point $\tau_E=0$: however for generic $\nu$, the derivative
${da\over d\tau_E}$ is not smooth so a Big-Bang type singularity
persists at $\tau_E=0$.

\section{Discussion}\label{sec:Disc}

Building on previous work on de Sitter extremal surfaces anchored at
the future boundary, we have studied no-boundary extremal surfaces in
slow-roll inflation models, with perturbations to no-boundary global
$dS$ preserving the spatial isometry. In pure de Sitter space
(Figure~\ref{fig1}) the Euclidean hemisphere gives a real area
equalling half de Sitter entropy \cite{Doi:2022iyj},
\cite{Narayan:2022afv}, while the timelike surfaces from the top
Lorentzian part give pure imaginary area (see (\ref{IRsurfAdSdS}),
(\ref{IRsurfAdSdS4}), (\ref{IRsurfAdSdS3})).  There is thus a clean
separation between the real and imaginary parts of these areas which
are best interpreted as time-entanglement or pseudo-entropy. A
heuristic Lewkowycz-Maldacena argument \cite{Narayan:2023zen}
executing the replica argument on $Z_{CFT}=\Psi_{dS}$ for maximal
no-boundary surfaces corroborates this interpretation since this is
now a replica on the Wavefunction regarded as an amplitude.
In the 
slow-roll case, we employ the explicit analytic expressions described
in \cite{Maldacena:2024uhs} for the slow-roll metric perturbations.
The presence of these perturbations (small wiggles around $dS$,
Figure~\ref{fig2}) mixes up everything so the no-boundary extremal
surface areas here have nontrivial real and imaginary pieces
overall. We evaluate the area integrals in the complex time-plane
defining appropriate contours designed to avoid extra poles that arise
at the complexification point due to the perturbations (similar
features arise in the evaluation of the Wavefunction of the
Universe). For the 4-dimensional slow-roll inflation case, the real
and imaginary finite corrections (\ref{sr4-area-final}) at leading
order in the slow-roll parameter match those in the semiclassical
expansion of the Wavefunction (or action). In 3-dimensional inflation,
we find that the real finite parts of the area (\ref{sr3-area-final})
agree with those in the Wavefunction (action): the imaginary finite
parts do not agree, perhaps reflecting the dual CFT on an even
dimensional sphere (with potential anomalies controlled by the
imaginary central charge in $dS_3/CFT_2$). Overall the maximal cosmic
brane creation probability controlled by the real parts of the
no-boundary extremal surface areas match the probability from the
semiclassical Wavefunction, corroborating the cosmic brane
interpretation in \cite{Narayan:2023zen}. The matching of these
maximal probabilities is perhaps expected since this is the maximal
amount of ``stuff'' in the space (in general dual conformal invariance
is broken but small corrections of the kind here do not wreck this
maximal probability). The slow-roll corrections decrease this maximal
probability (equal to half de Sitter entropy for pure $dS$), perhaps
reflecting the fact that small excitations in $dS$ decrease the
entropy. We have also studied no-boundary extremal surfaces in
Schwarzschild de Sitter spaces with small mass, with various overall
similarities, as well as other generic cosmologies.

Stepping back and looking at a broad-brush level, it is important to
understand the connections between the no-boundary extremal surface
areas and the Wavefunction of the Universe in greater detail
(mirroring connections between boundary entanglement entropy and
$Z_{CFT}$ in the exotic duals here): among other things this will shed
light on the underlying structure of these extremal surface areas in
$dS$ and cosmology more broadly. There is a rich history of such
studies in $AdS/CFT$ on extremal surface areas and the bulk partition
function (dual to entanglement entropy and the boundary partition
function) \cite{Ryu:2006bv,Ryu:2006ef,HRT,Rangamani:2016dms}.  One
such fact is that for maximal (IR) subregions in the $AdS$ black
hole/brane, the RT minimal surface dips all the way into the interior
and wraps the horizon so the finite part of entanglement entropy is
the black hole/brane entropy. This entropy can also be realized from
the action regarded as a partition function.  In the current context,
a similar observation is that the real finite part of the no-boundary
extremal surface in pure de Sitter is half $dS$ entropy, which also
arises from the real part of the Wavefunction.  It is natural to ask
if this correlation \cite{Narayan:2023zen} between the cosmic brane
creation probability (encoded by the real finite parts of the areas)
and $|\Psi_{dS}|^2$ continues away from de Sitter. An interesting
class of near $dS$ spaces comprises slow-roll inflation. Our studies
vindicate this correlation here.

As we have seen, we have found it necessary to define the area
integrals in the complex-time plane with appropriate contours required
to avoid extra poles at the complexification point. We have done this
in the minimal way, by simply evaluating the areas as separate
integrals in the Lorentzian and hemisphere regions, with the
regulating time contour Figure~\ref{fig3} serving to simply remove the
poles (and the details of the regularization are unimportant). The
integrals involve non-analytic expressions so one might expect that
changing the contour dramatically is not a reasonable thing to do in
these complex-valued integrals (and the Cauchy theorem cannot
apply). This can be corroborated further in the present case noting
that since the slow-roll case is near de Sitter we expect on
physical/geometric grounds that the time contour should also be near
that in de Sitter. In de Sitter the time contour encodes the IR
extremal surface shape in Figure~\ref{fig1} so it is natural that the
time contour in the slow-roll case encodes the wiggly shape for the IR
extremal surface in Figure~\ref{fig2}: this justifies the minimal time
contour Figure~\ref{fig3} that we have employed as quite reasonable
and natural here. Roughly the time contour dovetails with the
geometric picture of time in the semiclassical regimes we have here,
near pure de Sitter: indeed any time contour with large deviation from
the semiclassical geometry would perhaps be unphysical for the present
purposes. In general, these sorts of cosmological perturbations lead
to complex metrics that must be treated carefully (the complex nature
is not surprising since the no-boundary condition \cite{Hartle:1983ai}
is a regularity condition analogous to positive frequency not reality;
see comments in \cite{Maldacena:2024uhs} as well as
\cite{Maldacena:2002vr}).  Thus along the lines of the analysis here,
it would be interesting to understand if these sorts of no-boundary
extremal surface areas can be evaluated meaningfully and unambiguously
for complex metrics appearing in cosmology more generally, and whether
they give insights into complex metrics and the KSW criterion
\cite{Kontsevich:2021dmb,Witten:2021nzp} (see also
\cite{Hertog:2023vot}).

The classes of de Sitter perturbations we have studied here all
preserve the spherical isometry of the $S^3$ slices in global $dS_4$.
Thus all boundary Euclidean time equatorial plane slices are
equivalent, with the corresponding codim-2 no-boundary extremal
surface areas all equal. For more general inflationary perturbations with
inhomogeneities in particular, we expect that different boundary
Euclidean time slices are inequivalent and thus will lead to
inequivalent codim-2 extremal surfaces anchored at the future boundary
(an example of this inequivalence was already noted in
\cite{Narayan:2017xca}, \cite{Narayan:2022afv} for the extremal
surface areas in different slices of $dS$ in the static
coordinatization). It would be interesting to explore this further.

The extremal surfaces discussed here and in earlier work have been
studied directly and via analytic continuations from $AdS$ but they
are the analogs in some sense of the $AdS$ RT minimal surfaces on
constant time slices (which map to the boundary Euclidean time slices
stretching from the future boundary here). In that light, it is
natural to ask for covariant (HRT-like) formulations of these extremal
surfaces and the corresponding areas as pseudo-entropy. One might
imagine that the corresponding complex-time-plane contour analogs of
Figure~\ref{fig3} might then be expressed more covariantly.

Finally it would be interesting to understand the analyses here from
the point of view of the dual field theory. In a broad sense, $dS/CFT$
\cite{Strominger:2001pn,Witten:2001kn,Maldacena:2002vr} is a way to
organize de Sitter perturbations in the language of $AdS/CFT$ (and
analytic continuations) especially as formulated in
\cite{Maldacena:2002vr}, and in more recent studies
\eg\ \cite{Mata:2012bx,Ghosh:2014kba,Dey:2024zjx}.  In this light, it
would be interesting to understand the structure of inflationary
perturbations and the way they reflect in the no-boundary extremal
surface areas regarded as pseudo-entropy. Hopefully quantum information
ideas and techniques along these lines will help shed further light
on de Sitter holography and the emergence of time.

\vspace{4mm}

{\footnotesize \noindent {\bf Acknowledgements:}\ \ It is a pleasure
  to thank Pawel Caputa, Bartek Czech, Matt Headrick, Michal Heller,
  Alok Laddha, Shiraz Minwalla, Rob Myers, Suvrat Raju, Ronak Soni,
  Tadashi Takayanagi and Sandip Trivedi, on aspects of
  \cite{Narayan:2022afv,Narayan:2023zen}, and the present studies, as
  well as Ronak Soni and Tadashi Takayanagi for useful comments on a
  draft.  We thank the Organizers of the ``Quantum Information,
  Quantum Field Theory and Gravity'' Workshop, ICTS Bangalore, 2024,
  for hospitality while this work was nearing completion. This work is
  partially supported by a grant to CMI from the Infosys Foundation.
}


\appendix

\section{The basic setup on inflation}

\subsection{$dS_4$ slow roll}\label{App:dS4-sr-setup}

This pertains to sec.~\ref{dS4srnb} in the main text. We will be brief
in our review of pertinent points of inflation: general useful
reviews include \cite{Baumann:2009ds,Baumann:2014nda}, and
\cite{Hartle:1983ai}-\cite{Hertog:2023vot}.
The action for the 4-dim Einstein scalar theory (with $8\pi G_N=1$) is
\bea \label{dS4acslr}
& & I=\int d^4x \sqrt{g} \left(\frac{R}{2}-\frac{1}{2}\left(\nabla \phi\right)^2-V(\phi)\right)-\int d^3x\sqrt{h}K\,.
\eea
With $ds^2=-dt^2+a(t)^2 d\Omega_3^2,\ \phi=\phi(t)$, the equations of
motion become
\bea \label{dS4eoms-sr}
& & 3 H^2=3 \left(\frac{\dot{a}}{a}\right)^2=-\frac{3}{a^2}+\frac{1}{2}\dot{\phi}^2+V(\phi),\nn \\
& & \ddot{\phi}+3 H\dot{\phi}+V'(\phi)=0\,.
\eea
In the slow roll approximation, the inflaton kinetic term is
subdominant relative to the potential: for large 3-sphere size $a_r$
near the end of inflation, we ignore the $\frac{3}{a^2}$ curvature
term in (\ref{dS4eoms-sr}), which then gives
\bea \label{dS4eoms-sr-i}
& & 3 H^2 \sim V(\phi), \qquad 3 H\dot{\phi} \sim - V'(\phi)\,.
\eea
At ``horizon crossing'' $\phi=\phi_*$, we have\
$H_* \equiv H(\phi_*)=\frac{1}{a_*}=\frac{1}{a_r}e^{{\cal N}(\phi,\phi_*)}$\
and the leading $dS$ metric is (\ref{metric-atau-r}) with radius
$H_*^{-1}=l$, and
\be
l \equiv {1\over H_*}\,,\qquad H_*^2 \sim {V_*\over 3}\,,\qquad \tau = H_* t\,.
\ee
For the inflaton perturbation, the equation of motion alongwith the
slow-roll parameter $\epsilon$ gives
\be\label{sr-dS4-eom}
\phi=\phi_*+\varphi\,:\qquad
\partial_\tau^2 \varphi + 3 \tanh \tau \partial_\tau \varphi
+ 3\sqrt{2\epsilon}=0\,,\qquad\ \
\epsilon\equiv \frac{{{V_*}^\prime}^2}{2V_*^2}\,.
\ee
Solving this with\ $A=\frac{{V_*}^\prime}{H_*^2}=3\sqrt{2\epsilon}$\
gives 
\bea \label{soln-sr-dS4-eom}
& & \hskip -0.5in \varphi(\tau) =-\frac{1}{3} A \log (\cosh \tau )+\frac{1}{6} \text{sech}^2\tau  (2 A+3 c_1 \sinh \tau )+c_1 \tan
   ^{-1}\left(\tanh \frac{\tau }{2}\right)+c_2\,.
\eea
Imposing regularity at the no-boundary point $\tau=i\frac{\pi}{2}$
(via the $\varphi$-derivative) fixes $c_1=-\frac{2A}{3i}$. This then
gives
\bea \label{soln-sr-dS4-eom-i}
& & \varphi(\tau)= \frac{{V_*}^\prime}{V_*} \left(\frac{1+i \sinh \tau }{\cosh^2\tau }-\log (1-i \sinh \tau )-i{\pi\over 2}\right)=\frac{{V_*}^\prime}{V_*}\tilde{\varphi}(\tau)\,.
\eea
The constant ${\tilde c}_2$ (with $c_2=\frac{{V_*}^\prime}{V_*}{\tilde c}_2$
in (\ref{soln-sr-dS4-eom})) inside the bracket has been fixed to
$-i{\pi\over 2}$ by expanding at late times (large $\tau$), which gives
\bea \label{phitau-eq36}
\varphi(\tau)=\frac{V_*^{\prime}}{V_*}\tilde{\varphi}(\tau)=\frac{V_*^{\prime}}{V_*}\Biggl[-\tau+\log2+ i \frac{\pi}{2}+{\tilde c}_2+\ldots  \Biggr]\,.
\eea
This gives\ ${\tilde c}_2=-i \frac{\pi}{2}$\ to match with the
slow roll solution expanded around $\phi_*$ at large $\tau$\,:
\bea \label{phitau-eq36-large-tau}
& & \phi=\phi_*-\frac{V_*^{\prime}}{V_*}\log\left(\frac{a}{a_*}\right)= \phi_*-\frac{V_*^{\prime}}{V_*}\log\left(\frac{\cosh\tau}{H_* a_*}\right)
\sim \phi_*-\frac{V_*^{\prime}}{V_*}\left(\tau - \log 2\right)\,.
\eea
Putting all this together finally gives (\ref{dS4-varphitau}).
This completes our quick review, following \cite{Maldacena:2024uhs}.

\subsection{$dS_3$ slow roll}\label{App:dS3-sr-setup}

The overall setup for inflation is very similar so we will simply
list the central expressions used in sec.~\ref{dS3srnb}.\
The equations of motion (with $8\pi G=1$) are
\bea \label{dS3eoms-sr}
& &  H^2= \left(\frac{\dot{a}}{a}\right)^2=-\frac{1}{a^2}+\frac{1}{2}\dot{\phi}^2+V(\phi)\,,\nn \\
& & \ddot{\phi}+2 H\dot{\phi}+V'(\phi)=0\,.
\eea
In the slow roll approximation these give
\bea \label{dS3eoms-sr-i}
& &  H^2 \sim V(\phi), \qquad 2 H\dot{\phi} \sim - V'(\phi)\,,
\qquad\ra\qquad H_* \equiv H(\phi_*)\,,\quad  H_*^2 \sim V_*\,,
\eea
with the second set of expressions obtained when the 2-sphere size
crosses the horizon.

The inflaton equation (\ref{dS3eoms-sr}) expanding for the perturbation
as $\phi=\phi_*+\varphi$ becomes
\bea \label{sr-dS3-eom}
& & \partial_\tau^2 \varphi + 2 \tanh \tau \partial_\tau \varphi+\frac{{V_*}^\prime}{H_*^2}=0\,.
\eea 
With $\sqrt{2 \epsilon}=\frac{{V_*}^\prime}{H_*^2}=\frac{{V_*}^\prime}{V_*}$
this has the solution
\bea \label{soln-sr-dS3-eom}
& & \hskip -0.5in \varphi(\tau) =c_2+\tanh \tau \left(c_1-\frac{\sqrt{\epsilon} \ \tau}{\sqrt{2}}\right)\,.
\eea
Regularity at $\tau=i\frac{\pi}{2}$ fixes
$c_1=i\frac{\pi\sqrt{\epsilon}}{2\sqrt{2}}$\,. The constant $c_2$ is fixed
by matching with the slow roll solution expanded around $\phi_*$ at
late times (large $\tau$),
\bea \label{phitau-dS3-large-tau}
& & \phi=\phi_*-\frac{V_*^{\prime}}{V_*}\log\left(\frac{a}{a_*}\right)= \phi_*-\frac{V_*^{\prime}}{V_*}\left(\frac{\tau}{2}-\frac{\log2}{2}\right)\,.
\eea
This finally gives the inflaton profile as
\bea \label{soln-sr-dS3-eom-ii}
& & \varphi(\tau)= \sqrt{2\epsilon}\Biggl[\left(\frac{i\pi}{4}-\frac{\tau}{2}\right) \tanh\tau+\left(\frac{\log2}{2}-\frac{i\pi}{4}\right)\Biggr].
\eea

\section{Details: $dS_4$ slow-roll area}\label{App:dS4sr}

We describe details of the calculation of the no-boundary extremal
surface area for $dS_4$ slow roll inflation here.
The leading contribution to the area (\ref{sr4-area}) over the
contour (Figure~\ref{fig3}) is
\bea
S_0 &=& {\pi\,l^2\over 2G_4} \left(
\int_{y=1}^{y=\delta} {d(-y)\over 2\sqrt{y}}\ +\
\int_{\theta=\pi}^{\theta=0}
    {d(\delta\,e^{i\theta})\over 2\sqrt{-\delta\,e^{i\theta}}}\ +\ 
{1\over i} \int_\delta^{z_c} {dz\over 2\sqrt{z}} \right) \nn\\
&=& {\pi\,l^2\over 2G_4} \left(
-\sqrt{-z}\Big\vert_{-1}^{-\delta}\
+\ \sqrt{\delta} (-\sqrt{-e^{i\theta}})\Big\vert_{\theta=\pi}^{\theta=0}\
+\ {1\over i} \sqrt{z}\Big\vert_{\delta}^{z_c} \right) \nn\\
&=& {\pi\,l^2\over 2G_4} \left(
1-\sqrt{\delta}\ +\ \sqrt{\delta} (-i+1)\
+\ {1\over i} (\sqrt{z_c} - \sqrt{\delta}) \right) .
\eea
On the $[-1,0]$ leg we have redefined $z=-y$ which makes this piece
$\int_\delta^1 {dy\over 2\sqrt{y}}$\,. On the semicircle
we have $z=\delta e^{i\theta}$ as stated above and we then integrate
in terms of the variable $e^{i\theta}$. Of course here there is no
singularity at $z=0$ but we have written this explicitly above to
normalize this calculation with the slow-roll correction to follow. 
So we obtain simply\ (with $z_c\sim R_c^2$)\
$S = {\pi\,l^2\over 2G_4} ( 1 - i {R_c\over l} )$\,,\
after reinstating the dimensionful $dS_4$ scale $l$.

To calculate the $O(\epsilon)$ slow-roll correction, we first note
that the correction to the metric component $g_{rr}$ in (\ref{beta>(r)dS4})
is recast in the $z$-variable in (\ref{z-r}) for $z>0$ (\ie\ the
Lorentzian region $r>1$) as
\be
\beta_>(z)=\frac{i \left(3 \pi  (z+1)^3+\left(z^{3/2}+9 i z-12 \sqrt{z}-2 i\right) \left(-1-i \sqrt{z}\right)^3\right)}{6 z (z+1)^2}+\frac{(z+1) \log \left(1-i \sqrt{z}\right)}{z}\,,
\ee
and we obtain
\be
I_\epsilon^> = \int {\beta_>(z)\over 2i\sqrt{z}}\ dz\ =\
     {1\over 6 i \sqrt{z}} \left(2 - 3i\pi (1-z) - 7z
     - 6 (1-z) \log (1-i\sqrt{z}) \right) + {2/3\over 1-i\sqrt{z}}\,,
\ee
with $z>0$ in this expression.
It turns out happily that these integrals can be evaluated in Mathematica
(with some care in evaluating them as non-analytic complex objects):
the integrated expression can also be cross-checked manually of course.
To evaluate the slow-roll correction area integral in the hemisphere
region $r<1$, we continue to $z<0$: in this regard note that
\be\label{dS4-I>-I<}
\int_{-1}^{-\delta} {\beta_>(z)\over 2\sqrt{-z}} dz
= \int_1^\delta {\beta_>(-y)\over 2\sqrt{y}} d(-y)
= \int_\delta^1 {\beta_<(y)\over 2\sqrt{y}} dy
\equiv I_\epsilon^<(z)\,.
\ee
Thus it is adequate to analytically continue the integrated
expression $I_\epsilon^>(z)$ to the hemisphere as $z\ra -z$, and we obtain
the indefinite integral in the $O(\epsilon)$ piece here as
\be
I_\epsilon^<(z) = I_\epsilon^>(-z) = 
{1\over 6 \sqrt{z}} \left(-2 + 3i\pi (1+z) - 7z
+ 6 (1+z) \log (1+\sqrt{z}) \right) + {2/3\over 1+\sqrt{z}}\,,
\ee
with $z>0$ in this expression now.

Expanding $I_\epsilon^>$ in the $z>0$ part of the contour, we obtain
\be
z=z_c\ra\infty:\quad I_\epsilon^> = 1 + i {7\over 6} \sqrt{z_c}
-i\sqrt{z_c}\log\sqrt{z_c}\ ; \qquad
z=\delta :\quad
I_\epsilon^> = {2-3i\pi\over 6i \sqrt{\delta}} + {5\over 3}\ ,\quad
\ee
and we have regulated $z=0^+$ by $z=\delta>0$.
In obtaining the above, we have expanded $\log(1+z)\sim z$ for small
$z$ in the various expressions, and used $\log(-1)=i\pi,\ \log(\pm i)
= \pm {i\pi\over 2}$\,. This is important in order to cancel real
divergent terms for large $z_c$: as it stands we see that all the
divergent terms for large $z_c$ are pure imaginary.

Expanding $I_\epsilon^<$ in the $z<0$ part of the contour, we obtain
near $z=1$ (nbp) and $z=\delta$ (which is near $0^-$ in the earlier
variable)
\be
z=1:\quad I_\epsilon^< = {1\over 6} (-2 + 6i\pi - 7 + 12\log 2 )
+ {1\over 3}\ ,\qquad  
z=\delta :\quad
I_\epsilon^< = {-2+3i\pi\over 6 \sqrt{\delta}} + {5\over 3}\ .
\ee
Along the regulating semicircle around $z=0$ with small $z=\delta$,
examining the asymptotics of $I_\epsilon^>$, we obtain
\be
I_\epsilon^\theta\Big\vert_{\theta=\pi}^{\theta=0}
= \left({2-3i\pi\over 6\sqrt{-\delta e^{i\theta}}} \right)
   \Big\vert_{\theta=\pi}^{\theta=0}\ =\
   {2-3i\pi\over 6i\sqrt{\delta}} - {2-3i\pi\over 6\sqrt{\delta}}\ .
\ee
Note that near $z=0$, we have\ $\beta_>(z) = -{2-3i\pi\over 6z}$ as
the leading singular behaviour. Integrating this around the semicircle
explicitly recovers the above.

Adding up all the above contributions to the area (\ref{sr4-area})
finally gives (\ref{sr4-area-1}).

One may question whether the above calculation in the $z$-variable has
introduced subtleties due to the $\sqrt{-z}$ in various places.  It
turns out that the area can also be calculated directly in the form
(\ref{slowrollAreadS4dS3}) in the $r$-coordinate (\eg\ by evaluating
in Mathematica). Using $\beta_>(r)$ in (\ref{beta>(r)dS4}), we obtain
below the indefinite integrals $I^>(r)$ and $I^<(r)$ respectively\,:
\bea
-i\int {1+\epsilon\,\beta_>(r)\over \sqrt{r^2-1}}\ r\,dr &=&
-i\sqrt{r^2-1} -i \epsilon\, \frac{1}{6 r^2 \sqrt{r^2-1}} \Big[
\Big(3 i \pi  r^4-7 r^4-6 i \pi  r^2+5 r^2 \nn\\
&&\qquad +\ 4 \big(1+i \sqrt{r^2-1}\big)
+6 r^2 (r^2-2) \log \big(1-i \sqrt{r^2-1}\big)\Big)\Big]\,.\qquad
\eea
\bea
\int {1+\epsilon\,\beta_<(r)\over \sqrt{1-r^2}}\ r\,dr 
&=& -\sqrt{1-r^2} 
+ \epsilon \frac{1}{6 r^2 \sqrt{1-r^2}} \Big[ 
  \Big(3 i \pi  r^4-7 r^4-6 i \pi  r^2+5 r^2 \nn\\
&&\qquad\quad  +\ 4 \big(1-\sqrt{1-r^2}\big)
  +6 r^2 (r^2-2) \log \big(1+\sqrt{1-r^2}\big)\Big)
  \Big]\,.\qquad
\eea
In the second integral ($r<1$) above, we have obtained $\beta_<(r)$
by continuing $\beta_>(r)$ to $r<1$, which amounts to simply replacing
$-i\sqrt{r^2-1}$ by $\sqrt{1-r^2}$ in its four occurrences in the
expression (\ref{beta>(r)dS4}). Then we fix the overall sign of the
full integral so that the leading term matches the sign of the leading
term $\int_0^1 {rdr\over\sqrt{1-r^2}}$ in (\ref{slowrollAreadS4dS3})
which is the pure $dS_4$ hemisphere area in (\ref{IRsurfAdSdS}),
(\ref{IRsurfAdSdS4}). This is the form of the area integral in 
(\ref{sr4-area}) in the $r$-coordinate.
Evaluating these at the various limiting points gives
\bea
I^>(r)\Big\vert_1^{R_c/l} &=&  -i {R_c\over l} + \epsilon \left(1 
- i {R_c\over l} \log {R_c\over l} + i \frac{7}{6} {R_c\over l} \right)
- {5\over 3}\epsilon\ , \nn\\ [1mm]
I^<(r)\Big\vert_0^1 &=&  -{5\over 3}\epsilon - \Big[
-1+ \epsilon\,  \Big({7\over 6} - i \pi - \log 4\Big) \Big] ,
\eea
apart from singular terms at $r=1$ which can be removed by the
contour integral $I_\epsilon^\theta(r)$ around the regulating semicircle.
The above pieces alongwith the overall ${\pi\,l^2\over 2G_4}$ factor
add up to give (\ref{sr4-area-1}).

\section{Details: $dS_3$ slow-roll area}\label{App:dS3sr}

In this $dS_3$ case we have
\be\label{beta>(r)dS3-0}
\beta(r) = -\frac{1}{2} \left(r \frac{\partial \varphi}{\partial r}\right)^2
 + \frac{\varphi\, r^2}{1-r^2}\,.
\ee
Inputting the scalar profile (\ref{soln-sr-dS3-eom-ii}) in this case
and using $r=\cosh\tau$ gives
\bea\label{beta>(r)dS3}
\beta_>(r) &=&
\frac{1}{32 r^2 \left(r^2-1\right)}
\Big[    -4 r^4 (1+\log 16)+4 r^2-4 \log ^2\big(r+\sqrt{r^2-1}\big)\nn\\
  && \qquad\qquad\qquad +\, 4 \big(2 r \sqrt{r^2-1} \left(2 r^2-1\right)+i \pi \big) \log \big(r+\sqrt{r^2-1}\big) \nn\\
  && \qquad\qquad\qquad -\, 4 i \pi  \big(-2 r^4-r \sqrt{r^2-1} +2r^3 \sqrt{r^2-1} \big)+\pi ^2 \Big]\,.
\eea
In terms of the variable $z=r^2-1$ in (\ref{z-r}), this becomes
{\small\bea
\beta_>(z) &=& \frac{1}{32 z (z+1)}
\Big[ -4 (z+1)^2 (1+\log 16) + 4 (z+1) - 4 \log ^2\left(\sqrt{z}+\sqrt{z+1}\right)  \nn\\
  && \qquad\qquad\quad +\, 4 \left(2 \sqrt{z} \sqrt{z+1} (2 (z+1)-1)+i \pi \right) \log \left(\sqrt{z}+\sqrt{z+1}\right)  \nn\\
&& \qquad\qquad\quad -\, 4 i \pi  \left(2 \sqrt{z} (z+1)^{3/2}-2 (z+1)^2-\sqrt{z} \sqrt{z+1}\right) +\pi ^2 \Big]\,.
\eea }
Integrating gives
{\small\bea
I_\epsilon^> = \int {\beta_>(z)\over 2i \sqrt{z(1+z)}} dz &=&
  \frac{1}{32 \sqrt{z (z+1)}}
  \Big[  i \Big(8 i \pi  (z+1)-4 (2 z+1) \log ^2\left(\sqrt{z}+\sqrt{z+1}\right) \nn\\
&&   +\, \pi ^2 (2 z+1) + 4 i (2 \pi  z+\pi ) \log \left(\sqrt{z}+\sqrt{z+1}\right)-4 (z+1) \log 16\Big) \nn\\
&&  +\, 4 \sqrt{z} \sqrt{z+1} (2 \pi +i (1+\log 16)) \log \left(\sqrt{z}+\sqrt{z+1}\right) \Big]\,.
\eea }
Along the lines of the arguments in (\ref{dS4-I>-I<}) in the $dS_4$ case,
we have
\be\label{dS3-I>-I<}
\int_{-1}^{-\delta} {\beta_>(z)\over 2\sqrt{-z(1+z)}} dz
= \int_1^\delta {\beta_>(-y)\over 2\sqrt{y(1-y)}} d(-y)
= \int_\delta^1 {\beta_<(y)\over 2\sqrt{y(1-y)}} dy
\equiv I_\epsilon^<(z)\,.
\ee
Thus, analytically continuing the integrated expression $I_\epsilon^>$
to the hemisphere region ($r<1$) as $z\ra -z$ gives
{\small\bea
I_\epsilon^<(z) =  I_\epsilon^>(-z) &=&  \frac{1}{32 \sqrt{z (1-z)}}
  \Big[  \Big(8 i \pi  (1-z)-4 (-2 z+1) \log ^2\left(i\sqrt{z}+\sqrt{1-z}\right) \nn\\
&&   +\, \pi ^2 (-2 z+1) + 4 i (-2 \pi  z+\pi ) \log \left(i\sqrt{z}+\sqrt{1-z}\right)-4 (1-z) \log 16\Big) \nn\\
&&  +\, 4 \sqrt{z} \sqrt{1-z} (2 \pi +i (1+\log 16)) \log \left(i\sqrt{z}+\sqrt{1-z}\right) \Big]\,.
\eea }
Evaluating this at the various limiting points gives
\bea
z=z_c\ra\infty : &&
I_\epsilon^> = \frac{1}{16} i \left(-(\log z_c)^2+\log z_c+\pi ^2+4 i \pi
+ 4 (\log 2)^2-6 \log 2 \right) , \nn\\
z=\delta : &&
I_\epsilon^> = -\frac{\pi }{8}+\frac{i \left(8 i \pi +\pi ^2-4 \log 16\right)}{32 \sqrt{\delta}}\,.
\eea
Evaluating this now at $z=1$ (nbp) and $z=\delta$
(which is $0^-$ in the earlier variable) gives
\be
z=1 : I_\epsilon^< = {\pi\over 16} (2\pi i - 1 -\log 16)\,; \qquad
z=\delta : I_\epsilon^< = -\frac{\pi }{8}+\frac{ \left(8 i \pi +\pi ^2
  -4 \log (16)\right)}{32 \sqrt{\delta}}\,.
\ee
The contribution on the regulating semicircle is similar to that
in the $dS_4$ case and serves to cancel the singular pieces near $z=0$:
we obtain\
\be
I_\epsilon^\theta = 
-\frac{\left(8 i \pi +\pi ^2-4 \log 16\right)}{32 \sqrt{-\delta\,e^{i\theta}}}
\Big\vert_{\theta=\pi}^{\theta=0} =
-\frac{\left(8 i \pi +\pi ^2-4 \log 16\right)}{32 \sqrt{\delta}}\,
\Big({1\over i}-1\Big)\,.
\ee

Adding all these above contributions leads to (\ref{sr3-area-1}).
As in the $dS_4$ case, this area calculation can be done in the
$r$-coordinate also, using (\ref{beta>(r)dS3}), and its continuation
to the hemisphere.

\section{Inflation: on-shell action etc}\label{App:infl-Action}

We review aspects of evaluating the action for spaces with a
boundary: besides \cite{Hartle:1983ai,Bousso:1995cc,Bousso:1996au}, and
\cite{Maldacena:2024uhs}, some useful references include
\cite{Maldacena:2002vr,Hartle:2008ng,DiazDorronsoro:2017hti,Halliwell:2018ejl,Maldacena:2019cbz,Janssen:2020pii}
We consider the $d+1$-dim Einstein-scalar action 
\bea \label{dSd+1acslr}
& & I={1\over 16\pi G_{d+1}} \int d^{d+1}x \sqrt{g}
\left( R - \left(\del\phi\right)^2-2V(\phi)\right)
-{1\over 8\pi G_{d+1}}\int d^3x\sqrt{h}K\,.
\eea
With a minisuperspace-type metric ansatz as required for our analysis here
\be
ds^2 = -N^2(t) dt^2 + a^2(t) d\Omega_d^2\,,
\ee
and a $d+1$-split in the ADM formulation, we obtain the spatial
curvature and the extrinsic curvature as
\be
R^{(d)} = {d(d-1)\over a^2}\,,\qquad K_{ij}={1\over 2N} {\dot h}_{ij} =
{1\over 2N}\,2a{\dot a}\,s_{ij}\,,\quad K=s^{ij}K_{ij}={d\,{\dot a}\over a N}\,,
\ee
where $s_{ij}$ is the unit $S^d$ metric. Alongwith cancellations between
the bulk and boundary terms above, this gives the Lorentzian action
\bea
iI &=& {i\over 16\pi G_{d+1}} \int dt\, \Omega_d\, N\, a^d \left(
K_{ij} K^{ij} - K^2 + R^{(d)} - 2V + {1\over N^2} {\dot\phi}^2 \right) \nn\\
&=& {i\,\Omega_d \over 16\pi G_{d+1}} \int dt\, N\, a^d \left(
-d(d-1) {{\dot a}^2\over a^2 N^2} + d(d-1) {1\over a^2}
- 2V + {1\over N^2} {\dot\phi}^2 \right)\,.
\eea
Varying with $N$ leads to the Hamiltonian constraint with Lorentzian
signature:
\be
R^{(d)} - 2V - {1\over N^2} K_{ij} K^{ij} + {1\over N^2} K^2 -
{1\over N^2} {\dot\phi}^2\, =\, 0\, =\,
{d(d-1)\over a^2} + d(d-1) {{\dot a}^2\over a^2 N^2} - 2V 
- {1\over N^2} {\dot\phi}^2 \,.
\ee
To go Euclidean, we either take $N\ra iN$ or $t\ra it$ so
$-N^2dt^2\ra N^2dt^2$\,. This gives 
\be
-I_E = {\Omega_d \over 16\pi G_{d+1}} \int dt\, N\, a^d \left(
d(d-1) {{\dot a}^2\over a^2 N^2} + d(d-1) {1\over a^2}
- 2V - {1\over N^2} {\dot\phi}^2 \right)\,,
\ee
as the Euclidean action and the Hamiltonian constraint becomes
\be
{d(d-1)\over a^2} - 2V - d(d-1) {{\dot a}^2\over a^2 N^2}
+ {1\over N^2} {\dot\phi}^2 = 0\,.
\ee
For the metric in the Euclidean form on the right in
(\ref{metric-atau-r}), with $a$ taken as the time coordinate so
${\dot a}=1$, we then obtain 
\be
g_{aa} \equiv N^2\, =\, {{d(d-1)\over 2} - {1\over 2} (a\del_a\phi)^2
  \over {d(d-1)\over 2} - V\,a^2}\ .
\ee
Inputting the Hamiltonian constraint, we can evaluate the on-shell
action with the time coordinate taken as $a=t$.
We will now evaluate this for $dS_4$ and $dS_3$ slow-roll inflation
to $O(\epsilon)$ in the slow-roll parameter.

\bigskip

\noindent \underline{{\bf $dS_4$ slow-roll:}}\\
With $d=3$, inputting the Hamiltonian constraint, we obtain the
on-shell action
\be
-I_E = {6\Omega_3\over 16\pi G_4} \int da\,N\,a \left( {1\over N^2} + 1
- {a^2{\dot\phi}^2\over 6N^2} - {a^2V\over 3} \right)
= {3\pi\over 2G_4} \int da\,N\,a \left( 1 - {a^2V\over 3} \right)\,.
\ee
Thus the Lorentzian action taking $N^2\ra -N^2$ can be written with
the minus sign inside one radical and becomes
\be
iI = -{i\pi\over 2G_4} \int da\,a\, \sqrt{3-{1\over 2} (a\del_a\phi)^2}
\,\sqrt{a^2V-3}\,.
\ee
For pure $dS_4$, we have $V\equiv \Lambda={3\over l^2}$ and $\phi=const$,
giving the real part of the action as\
${\rm Re}(iI) = -{3i\pi\,l^2\over 2G_4}\int_0^1 dr\,r \sqrt{r^2-1}
= {\pi\,l^2\over 2G_4}$\,, \ie\ half $dS_4$ entropy as expected.

Incorporating the slow-roll correction and expanding to $O(\epsilon)$,
we obtain
\bea \label{If-vii}
& & iI = - {3i\pi\,l^2\over 2G_4} \int_0^{r_c} dr\,r\, \sqrt{r^2-1}\left[1+\epsilon\left(-\frac{1}{6}\left(r\partial_r {\tilde\varphi}(r)\right)^2+\frac{r^2 {\tilde\varphi}(r)}{r^2-1}\right)\right].
\eea
Breaking up the integral into the Euclidean part over $r\in [0,1]$ and
the Lorentzian part $r\in [1,r_c]$ with $r_c\equiv {R_c\over l}$\,, the
leading pure de Sitter term gives
\be
iI_0 = {\pi\,l^2\over 2G_4} \left[ 1
  - i\left( r_c^{3}-\frac{3}{2}r_c+O\left(\frac{1}{r_c}\right)\right)\right] .
\ee
For the $O(\epsilon)$ correction, we use the explicit expressions for
the inflaton profile and evaluate (\eg\ by defining an appropriate
contour to avoid the pole at $r=1$). Simplifying finally gives
\bea \label{I-final-i}
iI_\epsilon = {\pi\,l^2\over 2G_4}\,\epsilon\, \Biggl[
i r_c^3\left(\log r_c-\frac{1}{6}\right)+\frac{ir_c}{4}\left(6\log r_c-11\right)+\left(\log 4-\frac{7}{2}+i\pi\right)+O\left(\frac{1}{r_c}\right) \Biggr] .
\eea
This recovers the expressions in \cite{Maldacena:2024uhs}\ (see App.C.1
for the $O(\epsilon)$ terms), after noting ${V_*\over 3} = l^2$ and
reinstating $8\pi G_4$. 

\bigskip

\noindent \underline{{\bf $dS_3$ slow-roll:}}\\
With $d=2$, inputting the Hamiltonian constraint gives\
$I_E = -{2\Omega_2\over 8\pi G_4} \int da\,N\, ( 1 - a^2V )$,\ so the
on-shell Lorentzian action becomes
\be
iI = -{i\over G_3}\int da\,a\,
\sqrt{1-\frac{1}{2}(a\del_a\phi)^2}\,\sqrt{a^2V-1}\,.
\ee
Expanding to $O(\epsilon)$ and simplifying gives
\be
iI = - {il\over G_3} \int_0^{r_c} dr\, \sqrt{r^2-1}\left[1+\epsilon\left(-\frac{1}{2}\left(r\partial_r {\tilde\varphi}(r)\right)^2+\frac{r^2 {\tilde\varphi}(r)}{r^2-1}\right)\right] .
\ee
The leading term is the pure $dS_3$ action
\bea\label{srdS3action0}
iI_0 &=& - {il\over G_3} \left[\frac{i\pi}{4}
  + \frac{1}{2} \left(r_c\sqrt{r_c^2-1}-\log \left(r_c+\sqrt{r_c^2-1}\right)\right) \right] \nn\\
&=& {\pi\,l\over 4G_3} - {il\over G_3} \left( {r_c^2\over 2}
- {1\over 2}\log r_c - {1\over 2}\log 2 - {1\over 4} \right)
+ O\Big({1\over r_c}\Big)\,.
\eea
The subleading $O(\epsilon)$ terms can be evaluated as in the $dS_4$
case (\eg\ by an appropriate contour in the complex $r$-plane, avoiding
the pole at $r=1$). Simplifying, this finally gives
\bea
iI_\epsilon &=& {l\over 2G_3}\,\epsilon\,\Biggl[ -{\pi\over 16}(1+\log 16) \nn\\
&&\quad  +\ \frac{i}{16} \left(  4 z_c\log z_c-2z_c 
  +  (\log z_c)^2+\log z_c 
  + \pi ^2+1 - 4 (\log 2)^2-2 \log 2 \right) \Biggr] . \quad
\eea
Using $z_c=r_c^2-1$ and expanding gives
\bea\label{srdS3action-epsilon}
iI_\epsilon &=& {l\over 2G_3}\,\epsilon\,\Biggl[ -{\pi\over 16}(1+\log 16) \nn\\
&& \ +\ {i\over 16} \left( 8 r_c^2\log r_c -2r_c^2
  + 4 (\log r_c)^2 - 6 \log r_c 
  + \pi ^2-1 - 4 (\log 2)^2 - 2 \log 2 \right)  \Biggr] .\qquad
\eea

\end{document}